\documentclass[a4paper,10pt,oneside]{article}

\usepackage{tgpagella}
\usepackage[T1]{fontenc}
\usepackage{latexsym}
\usepackage{amssymb}
\usepackage{amsmath}
\usepackage{float}
\usepackage{mathtools}
\DeclarePairedDelimiter\ceil{\lceil}{\rceil}
\DeclarePairedDelimiter\floor{\lfloor}{\rfloor}

\usepackage{theorem}

\usepackage{hyperref}

\usepackage{pgf}
\usepackage{xxcolor}

\usepackage{natbib}

\theoremheaderfont{\scshape}
\theorembodyfont{\rmfamily}

\bibpunct[, ]{(}{)}{,}{a}{}{;}

\def\qed{\relax
   \ifmmode
    ~\hfill\Box
   \else
    \unskip\nobreak ~\hfill$\square$%
   \fi \par}

\newcommand{\sep}{$\cdot$ }

\begin{document}
	\title{{\scriptsize -- final manuscript accepted for publication in Logique
			et Analyse --}\\
	On the Nature of Discrete Space-Time \\ 
	\large Part 1:  The distance formula, relativistic time dilation \\ and length contraction in discrete space-time}
    
	\author{DAVID CROUSE and JOSEPH SKUFCA}
	\date{}
	\maketitle

\begin{abstract}
	In this work, the relativistic phenomena of Lorentz-Fitzgerald contraction and time dilation are derived using a modified distance formula that is appropriate for discrete space.  This new distance formula is different than the Pythagorean theorem but converges to it for distances large relative to the Planck length.  First, four candidate formulas developed by different people over the last 70 years is discussed.  Three of the formulas are shown to be identical for conditions that best describe discrete space.  It is shown that this new distance formula is valid for all size-scales -- from the Planck length upwards -- and solves two major problems historically associated with the discrete space-time (DST) model. One problem it solves is the widely believed anisotropic nature of most discrete space models.  Just as commonly believed is the second problem -- the incompatibility of DST's concept of an \textit{immutable} atom of space and the length contraction of this atom required by special relativity.  The new formula for distance in DST solves this problem. It is shown that length contraction of the atom of space \textit{does not occur} for any relative velocity of two reference frames.  It is also shown that time dilation of the atom of time does not occur.  Also discussed is the possibility of any object being able to travel at the speed of light for specific temporal durations given by an equation derived in this work. Also discussed is a method to empirical verify the discreteness of space by studying any observed anomalies in the motion of astronomical bodies, such as differences in the bodies' inertial masses and gravitational masses. The importance of the new distance formula for causal set theory and other theories of quantum gravity is also discussed. 

\end{abstract}

\section{\label{sec:intro}Introduction}
Can spatial distances be divided into ever smaller segments, or is there a limit to which any spatial distance can be subdivided?  Likewise with time: is time continuous or is there a smallest temporal duration? In other words, does time flow as a continuous stream or as a discrete series of snapshots? These and similar questions were first posed by Greek and medieval philosophers Parmenides, Zeno of Elea \citep[9]{Hagar2014} and Maimonides (\citet{Maimonides1190}).  Since then, the debate on the idea that space and time are atomized, or discretized, has waxed and waned. In the modern age, Werner Heisenberg had a continued interest in discrete space, along with his contemporaries:  Arthur March, Henry Flint, Arthur Ruark and others \citep[70, 99-103]{Hagar2014}. Over the last 40 years there has been a resurgence of interest in this concept due to the emergence of new physical theories, including loop quantum gravity (\citet{Pullin2011, Rovelli2003, Collins2004}) and causal set theory (\citet{Henson2008, Sorkin2005}), along with recent work in mathematical physics (\citet{Hagar2014, Finkelstein1969}) and pure mathematics (\citet{Forrest1995, Weyl1949, VanBendegem1987, VanBendegem1995}).  

Our interest in discrete space-time (DST) started from a much different topic: a simple exercise where we examined a possible explanation for the observed inertial anomalies of astronomical bodies -- anomalies that are normally attributed to dark matter and dark energy (\citet{Crouse2016a}). In that work, we treated John A. Wheeler's quantum foam that ostensibly pervades all space (\citet{Wheeler1957}) as a solid-state material, and then studied the motion of particles and astronomical bodies traveling within this material. The important thing we realized was that Wheeler's assumption of the foam being a random distribution of particles (each of Planck mass $m_p=2.18 \times 10^{-8}$ kg) with an \textit{average} particle-to-particle spacing of the Planck length ($l_p=1.62 \times 10^{-35}$ m) cannot be correct. Instead, if space is discretized in spatial units of $l_p$, then the foam must be a crystal with a particle-to-particle spacing of \textit{exactly} $l_p$.  We then used simple crystal physics (with a spatially periodic gravitational potential energy profile) to calculate the energy bands and effective inertial masses of particles traveling within this universe-wide gravity crystal. As discussed in Section \ref{sec:Crystal}, we found that the crystal supports particle behavior that mimics the effects of dark matter and dark energy. Observing and quantifying such behavior can provide a realizable way to verify that space is discrete.  

However, while we were investigating the use of the DST model to justify the ordering of the quantum foam, we came across something even more interesting:  the unresolved problems associated with DST and especially Hermann Weyl's 1949 tile argument -- see (\citet{Hagar2014}) and \citep[p. 43]{Weyl1949}.  Weyl purported that this argument required one to either accept the Pythagorean theorem \footnote{More specifically, the calculation of distances in flat-space using the Pythagorean theorem. \label{footnote 1}} or the discretization of space, but not both.  Since then, Weyl's argument has proven difficult to refute, with no significant progress having been made to counter it until Jean Paul Van Bendegem's 1987 paper on the topic (\citet{VanBendegem1987}).  We were amazed and perplexed by this. But above all, we thought that for humanity's oldest mathematical theorem to still be in doubt was an embarrassment for us all, and needed to be resolved posthaste.  We also realized that a solution to the problem with the Pythagorean theorem$^{\ref{footnote 1}}$ would resolve many of the simple problems with DST, and any problems that remained would be extraordinarily interesting and worthy of additional study.  Thus, in this paper we solve the problems:  conservation of the invariant/immutable nature of the atom of space (the hodon $\chi$) and of the atom of time (the chronon $\tau$) (i.e., $\chi$ and $\tau$ should not experience relativistic length contraction and time dilation respectively), maintaining isotropy in DST, causality issues with DST, and the calculation of distances at all size-scales (from the Planck scale to macro-geometric scales) using a single general formula.  The only issue that remains with DST is the conservation of energy-momentum, and as will be discussed in Section \ref{sec:Discussion}, this sole remaining issue is rich with implications on our conceptions of motion, energy, and mass.  

We were also interested in this topic for one particular purely philosophical reason: to see if there is any chance to resuscitate Henri Bergson's once stellar reputation as \textit{the} authority on the subject of time -- a renown that diminished to the point of obscurity after Albert Einstein's introduction of the special theory of relativity (SR).  Bergson strongly opposed equating scientists' ``measured time'' with what he believed to be a more important and human-experienced time that he called ``real time'', ``psychological time'', ``lived time'', duration, or simply \textit{Time} (\citet{Canales2015}).  Even though he later came to accept the time dilation of physicists' time, he thought that his \textit{Time} is ``not altered according to the velocity of a system'' \citep[p. 25]{Canales2015}.  To further this view, he quipped  ``We shall have to find another way of not aging'' \citep[p. 11]{Canales2015} \citep[p. 77]{Bergson1926}.  Albert Einstein, however, famously stated that ``the time of the philosophers does not exist'' \citep[p. 19]{Canales2015}.  Thus, we were interested to see if there is some compromise that DST may allow between these two opposing views.  Does DST contain some aspect of time that is not dilated and is universally experienced by all entities?  Unfortunately for Bergson, what we found in this work is that the only aspect of time that is not affected by velocity is the chronon.  And while the chronon (determined later in this work to be $\tau=2\tau_p=1.08 \times 10^{-43} s$ with $\tau_p$ being the Planck time) does not experience time dilation, it is far removed from any human-experienced time, and ironically, it is the physicists' time that is the human-experienced time.

Having resolved all but one of the commonly cited problems with DST, one way to view this paper is simply as a strong defense of the DST model against the continuous model.  However, we view this paper as more than just this. Having a correct expression of the formula to calculate distances in space is important for many reasons, with different reasons being important for different communities of scientists and philosophers. For the relativists, this paper will be important for the formulas describing the distance formula, Lorentz factor, time dilation, and length contraction in DST. For the quantum gravity, astronomy, and cosmology scientists, this paper will be important for how it imposes order upon Wheeler's quantum foam, as well as its impact on quantum loop gravity and causal set theory. For the mathematicians, it will be important because it defines a new pseudo-metric that violates the triangle inequality theorem in interesting ways. I know that as a materials engineer, I am interested in further investigating the universe-wide gravity crystal, how it forms, the growth dynamics in its bulk and at its facets, the properties of defects in the crystal, and the drift and diffusion of mobile particles (e.g., black holes) within the crystal.  And finally for philosophers, this paper will be important for its defense/elevation of the DST model and its philosophical implications on space, time, and motion.  However, we are confident that we are overlooking many other fields of study for which this most basic tool, the distance formula, is important now and increasingly so in the future as humanity probes, studies and exploits phenomena that occur at ever smaller size-scales.

This paper is organized as follows. In Section \ref{sec:Distance}, four different versions of the distance formula applicable to discrete space are discussed - those proposed by Hermann Weyl (\citet{Weyl1949}), Jean Paul Van Bendegem (\citet{VanBendegem1987}), Peter Forrest (\citet{Forrest1995}) and David Crouse (\citet{Crouse2016b}).  It is discussed how three of these formulas, with some slight modifications and reinterpretations, are identical for conditions that best describe DST, and how this formula differs from the Pythagorean theorem for small size-scales but converges to it for any distance significantly larger than $l_p$.  In Section \ref{sec:TimeLength}, the standard derivations of time dilation and length contraction found in any textbook on SR are performed, with the only change being that the modified distance formula is used in the derivations instead of the Pythagorean theorem.  It is in this section that we develop a new Lorentz factor $\gamma$ that eliminates the contraction of the hodon and the dilation of the chronon that are predicted by conventional SR.  An interesting consequence of these results is discussed in Section \ref{sec:LightSpeed}, namely, how SR in DST allows objects to travel at the speed of light over specific temporal durations.  In Section \ref{sec:Crystal}, DST's impact on Wheeler's quantum foam is discussed.  Additional items are discussed in Section \ref{sec:Discussion}, including causality, conservation of energy and momentum, Mach's principle, gravity's impact on the DST model, and a discussion of the importance of our DST model to causal set theory.

\section{\label{sec:Distance}Proposed Distance Formulas for Discrete Space}

\subsection{Hermann Weyl's Distance Formula}
\begin{figure}[H]
 	\centering\includegraphics[width=5cm]{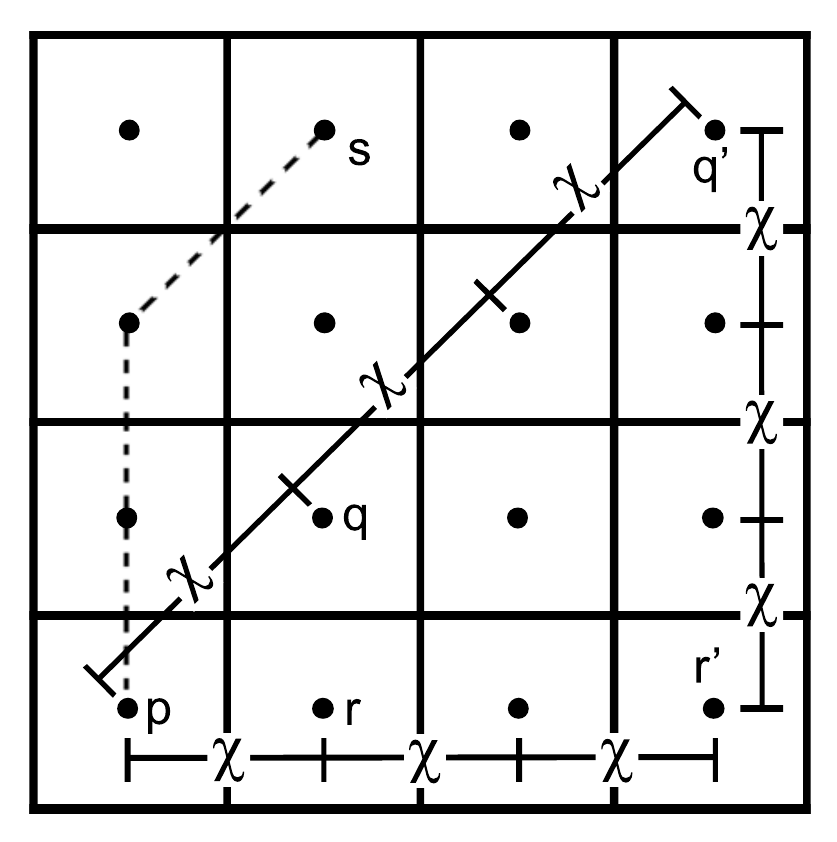}
 	\caption{\textbf{Grid and Solid Lines:} The Weyl construction that includes the \textit{a priori} existence of a lattice.  All distances from the center of one tile to the center of any neighboring tile must be $\chi$. Thus the length of the diagonal is equal to the length of the side of the square ($\overline{pq}=\overline{pr}=\chi$) regardless of the size of the square (e.g., $\overline{pq'}=\overline{pr'}=3\chi$).   \textbf{Dashed Line:} The distance $\overline{ps}$ is the shortest path composed of multiple jumps, with each jump being of extent $\chi$; thus $\overline{ps}=3\chi$. The path shown is only one of several that yield $\overline{ps}=3\chi$.}.
 	\label{fig:Fig1}
 \end{figure}

In 1949, Hermann Weyl constructed his famous Weyl-tile argument against discrete space \citep[p. 43]{Weyl1949}.  The argument starts by modeling DST as a fixed grid of identical ``tiles" with a tile-to-tile spacing of $\chi$ (Fig. \ref{fig:Fig1}), with $\chi$ being nature's ostensibly minimum length, i.e. the hodon, which we determine later in this paper to be $\chi = 2l_p=3.24 \times 10^{-35}$ m. Weyl then argued that the length of the hypotenuse of an isosceles right triangle (i.e., $c$) is equal to the length of the side of the triangle (i.e., $a$).  The important point in his argument is that $c$ and $a$ are equal \textit{regardless of the size of the triangle}, and because we measure (in actuality) the length of the hypotenuse of any such triangle to be $\sqrt{2}$ times the length of its side, space must not be discrete.  Even though Weyl did not do so, one can easily develop a distance formula based on his construction: 

\begin{equation}\label{Weyl}
d = m \chi.
\end{equation}

\noindent where $d$ is the distance between two points, and $m$ is the integer number of steps of the \textit{shortest} path, with each step going from the center of one tile to the center of an adjacent tile either at its side or diagonal to it. An example is shown by the dashed line in Fig. \ref{fig:Fig1} from point $p$ to $s$ with a length $\overline{ps}=3\chi$.

Even though Eq. \eqref{Weyl} leads to demonstrably incorrect results for macro-geometric distances, Weyl was the first person to suggest any modification of the distance formula given by the Pythagorean theorem \footnote{In this paper we are not considering a curved space structure promoted by general relativity. As discussed in Section \ref{sec:SpaceTimeAtom}, in the model developed in this paper there is no \textit{a priori} existing space to curve!}.  It went unrecognized at that time and up to now, that even though Weyl's tile argument created one problem with the DST model (i.e., disagreement with the Pythagorean theorem), it solved two other problems with the model.  Namely, Weyl's result of the length of the hypotenuse being $\chi$ for the smallest sized triangle (i.e., $a=b=\chi$) solved the problems of length contraction of the hodon and time dilation of the chronon (see Section \ref{sec:TimeLength}). Even if this had been recognized, it would have produced little consolation, since Weyl's argument seemed to be otherwise so damning for the DST model. For the next 35 years the situation changed little. Then in 1987, Jean Paul Van Bendegem pointed us all in the right direction towards a proper resolution of the Weyl-tile argument (\citet{VanBendegem1987}).

\subsection{Jean Paul Van Bendegem's Distance Formula}
\begin{figure}[H]
	\centering\includegraphics[width=7cm]{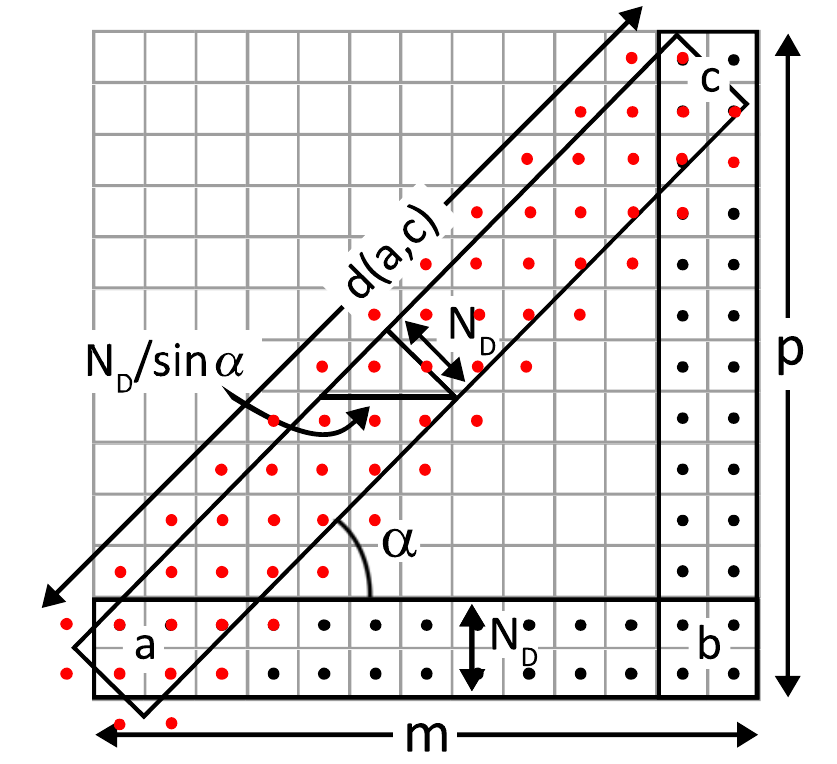}
	\caption{Van Bendegem's method involves a grid and lines with widths $N_D$.  The distance from point $a$ to point $c$ is the sum of all the squares (with red dots) within or touched by the line, divided by $N_D$.}
	\label{fig:Fig2}
\end{figure}

In a 1987 paper on Zeno's paradoxes and the Weyl-tile argument, Van Bendegem made the four following assumptions about lines in discrete space \citep[pp. 296-298]{VanBendegem1987}:

\begin{enumerate}
  \item In a discrete geometry, all lines must have a constant nonzero width (of integer $N_D$).
  \item A line consists of all the squares that are touched during the act of drawing the line (see Fig. \ref{fig:Fig2}).
  \item The size of the squares is small compared to the macroscopic width of the lines.
  \item The length of a line is the sum of the squares constituting that line, modulo the width.
\end{enumerate}

\noindent with $N_D$ being an integer, and the actual width of a line in units of length being $N_D \chi$, with $\chi$ being the minimum length as before. In his later works (\citet{VanBendegem1995, VanBendegem2017}), Van Bendegem dropped the necessity of Assumption 3 and even considered the case with $m=N_D=1$ as resolving the Weyl-tile argument.  Applying this procedure to \textit{straight} lines, say $m$ rectangles long and $N_D$ rectangles wide, one obtains the expected result (\citet{VanBendegem1987}):

\begin{equation}\label{BendegemStraightLine}
  Length = m \cdot N_D \left ( \mbox{div } N_D \right )=m.
\end{equation}

\noindent Thus, the length of the base of the triangle shown in Fig. \ref{fig:Fig2} is $m$ and the height is $p$.  Now concerning the hypotenuse, Van Bendegem states that ``the hypotenuse can be considered as a vertical pile of $p$ layers'' (\citet{VanBendegem1987}).  This leads to the equation, given in (\citet{VanBendegem1987}) and more clearly in (\citet{VanBendegem2017}):

\begin{equation}\label{BendegemGeneralA}
d(a,c) = p \cdot \left \lfloor \frac{N_D}{sin \alpha} \right \rfloor div \left ( N_D \right ).
\end{equation}

\noindent Note that Eq. \eqref{BendegemGeneralA} is not quite the same as Assumption 2, but is better because it removes any ambiguity as to which rectangles to include in the sum used to calculate $d(a,c)$. Importantly, in (\citet{VanBendegem1995}) Van Bendegem expressed Eq. \eqref{BendegemGeneralA} in a slightly different way: he placed the factor $p$ in the floor operation in Eq. \eqref{BendegemGeneralA}, resulting in:

\begin{equation}\label{BendegemGeneralB}
d(a,c) = \left \lfloor p \cdot \frac{N_D}{sin \alpha} \right \rfloor div \left ( N_D \right ).
\end{equation}

\noindent This change is significant because it is a step towards connecting Van Bendegem's approach to the approach taken by Crouse in which he rejected the \textit{a priori} existence of a grid (\citet{Crouse2016b}).  The next step involves letting $N_D=1$ in Eq. \eqref{BendegemGeneralB} (which seems most appropriate for discrete space). Equation \eqref{BendegemGeneralB} then becomes

\begin{equation}\label{BendegemGeneralC}
d(a,c) = \left \lfloor \frac{p}{sin \alpha} \right \rfloor = \left \lfloor \sqrt{m^2+p^2} \right \rfloor .
\end{equation}

\begin{figure}[H]
    \centering\includegraphics[width=7cm]{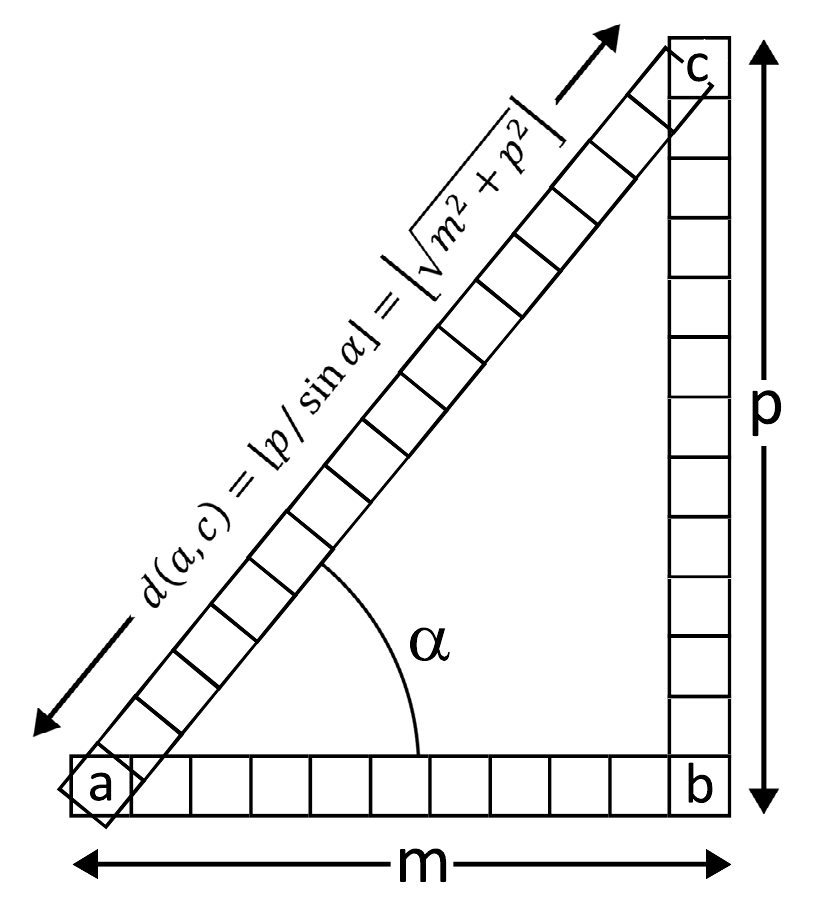}
    \caption{An alternative form of Van Bendegem's distance formula (i.e., Eq. \eqref{BendegemGeneralC} that uses $N_D=1$) suggests that the grid does not have a preferred direction. Isotropy is maintained in this model and the distances Eq. \eqref{BendegemGeneralC} predicts converge to those given by the Pythagorean theorem for large distances.}
    \label{fig:Fig3}
\end{figure}

It is seen that Eqs. \eqref{BendegemGeneralB} and \eqref{BendegemGeneralC} (but not Eq. \eqref{BendegemGeneralA}) converge to the Pythagorean theorem for large $m$ or large $p$.  Importantly though, one can interpret Eq. \eqref{BendegemGeneralC} somewhat differently, namely, interpreting it as the consequence of $d(a,c)$ being the number of complete rectangles (including the rectangle centered about $a$) included along a \textit{tilted column} along the hypotenuse (see Fig. \ref{fig:Fig3}).  The benefit of this modification is that it suggests a solution to the supposed anisotropy problem with DST:  when an entity travels from $a$ to $b$, the grid manifests locally as appropriate to that direction of travel, when traveling from $a$ to $c$, the grid manifests locally as appropriate to this different direction of travel.  Namely, the grid does not exist \textit{a priori} from the perspective of a  particle.  One may then be tempted to say that the grid exists \textit{a posteriori}, coming into existence and remaining in existence only \textit{locally} (i.e., within the immediate neighborhood of the particle). However, if the grid is so ephemeral (along with its inability to affect a particle's direction of travel), it is logical to want to discard the grid entirely; this was done by Crouse in (\citet{Crouse2016b}).  However, before describing Crouse's grid-less approach, we discuss below another important grid-based method developed by Peter Forrest (\citet{Forrest1995}).

\subsection{\label{sec:Forrest}Peter Forrest's Distance Formula}
\begin{figure}[H]
	\centering\includegraphics[width=7cm]{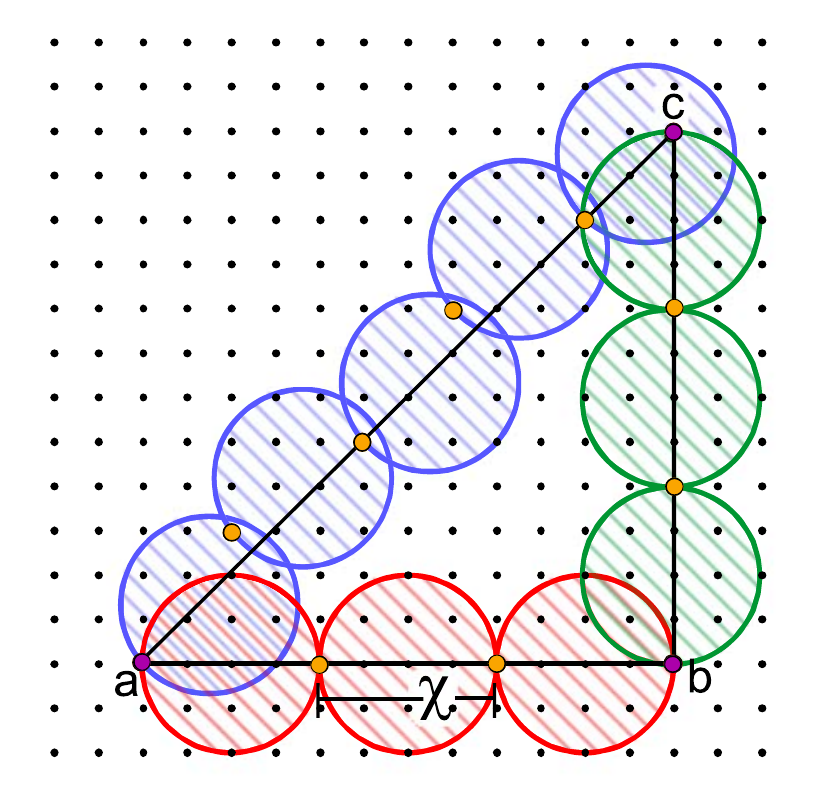}
	\caption{In Forrest's approach, a grid is constructed (black dots) and the BoAs/links (with $m=2$ in this example) are shown along the base (dashed red circles), height (dashed green circles) and hypotenuse (dashed blue circles).  Particular grid-points that ``connect'' two link-points are shown in orange.}
	\label{fig:Fig4}
\end{figure}

In 1995, Peter Forrest (\citet{Forrest1995}) sought to develop a distance formula appropriate for discrete space that uses ``only a single dyadic relation of \textit{adjacency}''; he pointedly rejected Van Bendegem's approach that he thought did not ``define distance in terms of adjacency''. Forrest's approach is interesting and rests on an intriguing, but we think problematic definition of adjacency. Forrest states that the distance between two points $p$ and $q$ is the number of links in the chain, with the first link containing $p$ and the last link containing $q$ (as shown in Fig. \ref{fig:Fig4}).  Of the multitude of possible chains connecting $p$ and $q$, the appropriate one for the determination/assignment of distance is the chain that has ``the least number of links \dots, the first of which is [contains] $p$ and the last of which is [contains] $q$''.  A ``link'' in Forrest's model is a collection of points that he states are all ``adjacent'' to each other, with any two points $\left \langle u,v \right \rangle$ and $\left \langle x,y \right \rangle $ in $\textbf{E}_{2,m}$ being adjacent if they satisfy the equation

\begin{equation}\label{Forrest}
\left ( u-x \right )^2 + \left ( v-y \right )^2 \leq m^2,
\end{equation}

\noindent where $m$ is a scale factor for which Forrest proposes a value of $10^{30}$ as being appropriate for the real space in which we live.  Two links are said to be ``connected'' if they have one or more points $\left \langle u,v \right \rangle$ in common. He calls the links ``balls of adjacency'' (BoAs), and each BoA may contain a large number of points all ``adjacent'' to each other.  Again, the distance from $p$ to $q$ is the number of links in the shortest chain connecting $p$ to $q$. 

A note is in necessary concerning the classification and labeling of points in Forrest's method.  There are two types of points in Forrest's approach:  grid points and link points.  Grid-points are single identifiable and unique entities (the black dots in Fig. \ref{fig:Fig4}) that can be specified by their $x$ and $y$ components within angled brackets $\left \langle u,v \right \rangle$.  Link-points, or just links or BoAs, contain a multitude of grid-points, examples of which are the blue, red and green circles in Fig. \ref{fig:Fig4}.  Confusion can occur with this nomenclature however, since we will sometimes refer to a point as both a grid-point and a link-point; by this we mean a grid-point, say $a$, and all the other grid-points adjacent to $a$ (according to Eq. \eqref{Forrest}), or in other words, within one of the BoAs/links associated with grid-point $a$.  Additionally, in Section \ref{subsec:Triangle} we need to make use of a labeling system for link-points that specifies their $x$ and $y$ components in \textit{link-points}.  What this means is that a link-point $e$ is labeled as $e=[g,h]$ using square brackets where $g$ is the smallest number of link-points one has to translate in the $x$ direction, followed by the smallest $h$ link-point translation in the $y$ direction to arrive at link-point $e$ after starting from the link-point origin $[0,0]$.  

Two benefits of Forrest's approach are: the anisotropic nature of the grid can be minimized by letting $m$ be large, and the distances it predicts for large triangles converge to those predicted by the Pythagorean theorem.  However, one major problem with Forrest's method is that, as formulated, the distances it calculates are generally at least one integer larger than those predicted by Eq. \eqref{BendegemGeneralC}.  It will be shown in the next section that because of this, Forrest's method leads to one very important non-physical result.

\begin{figure}[H]
    \centering\includegraphics[width=10.0cm]{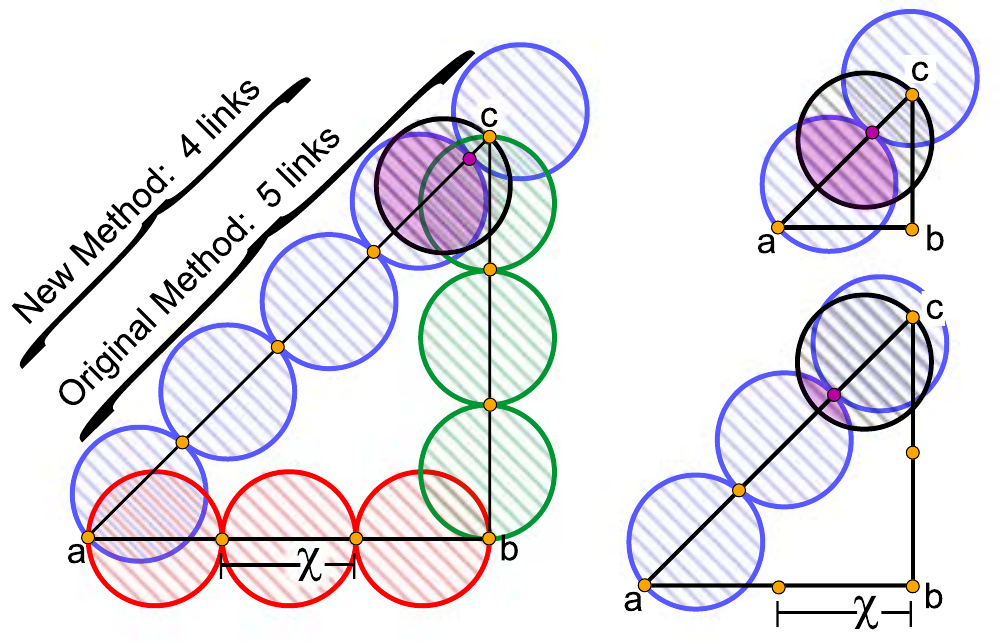}
    \caption{\textbf{Left:} A schematic showing a different way of interpreting Forrest's approach.  First, let the spacing between the grid-points go to zero, i.e., the grid is infinitely dense.  The grid then disappears, and with it, the anisotropy of Forrest's construction.  Also, we rethink when the ``chain'' (in blue) has reached the end point $c$.  We do not require the last link in the chain to contain the grid-point $c$ but only partially overlap $c$'s BoA (the black/gray shaded circle). The grid-points common to both BoA's (i.e., those of $c$ and of the last link in the chain) are shaded in purple. \textbf{Right:} Two other triangles with base and height lengths of $\chi$ and $2\chi$, showing both the original and new methods of determining the lengths of the hypotenuses.  These are shown to assist the reader in studying the model's adherence to the triangle inequality theorem. }
    \label{fig:Fig5}
\end{figure}

One way to fix this discrepancy is to first let $m \rightarrow \infty$ and then construct Forrest's BoAs slightly differently, as shown in in Fig. \ref{fig:Fig5}. This difference exploits the ambiguity in the placement and orientation of $c$'s BoA (the link colored in black in Fig. \ref{fig:Fig5}).  The difference is, when calculating $d(a,c)$, (where $a$ and $c$ are link-points), place and orient grid-point $c$'s BoA such that grid-point $c$ is at one side of a BoA, and the opposite side of this BoA is oriented towards grid-point $a$, as shown in Fig. \ref{fig:Fig5}.  In this configuration, the last link in the shortest chain need only contain one grid-point in $c$'s BoA rather than grid-point $c$ itself. The grid-points that are in both the last link and $c$'s BoA are shaded in purple in Fig. \ref{fig:Fig5}.  This approach eliminates the anisotropy that existed when $m$ is a finite number and yields distances matching those predicted by Eq. \eqref{BendegemGeneralC}.  However, even though this modification results in agreement between Van Bendegem's and Forrest's approaches, one would be justified in feeling uneasy with the foundations of these approaches that rely on grids and/or balls of adjacency.  As described next, in a 2016 paper (\citet{Crouse2016b}) Crouse derived Eq. \eqref{BendegemGeneralC} in a way that did not involve the \textit{a priori} existence of a grid, or BoAs with interior structure/features (i.e., grid-points).

\subsection{David Crouse's Distance Formula}
In 2016, upon completing the analysis of the impact of the DST model on Wheeler's quantum foam (\cite{Crouse2016a}), Crouse considered how one could fix the problems with DST.  Being at the time unburdened by any knowledge of previous work on DST (besides Weyl's work), or the fact that logical positivism (LP) and the concept of non-absolute space have fallen out of favor with physicists, mathematicians and philosophers, it took little time for us to derive Eq. \eqref{BendegemGeneralC} using a different approach.  We knew from the offset that our approach solved the anisotropy problem that is said to exist with DST, and it took little longer for us to realize that it also solved the problems of length contraction of $\chi$ and time dilation of $\tau$.  The calculation involved two steps: the derivations of $\chi$ and $\tau$, and the derivation of the new distance formula Eq. \eqref{BendegemGeneralC} or \eqref{LeopoldB}.  Both steps are described in detail in this section.

\subsubsection{\label{sec:SpaceTimeAtom}The Atoms of Space and Time}
We start the derivations of the atoms of space and time by invoking a restrictive form of LP and non-absolute space (NA-space).  While there are a lot of problems with LP in general, (as evidenced by some of Ernst Mach's beliefs \citep[p. 232]{Eberhard1989}), I am an strong believer in a \textit{restrictive form} of LP.  This form of LP is so restrictive that is not of much use as a general philosophical school of thought, but fortuitously, it is exactly what we need to determine the atoms of space and time, and it is so conservative that one can have a high degree of confidence with the approach.  Specifically, the form of LP used in this paper involves a stringent test for a thing's exclusion from reality:  a thing is excluded from reality only if it cannot be measured by (or make its presence known to): anybody (i.e., any measurement system), anywhere, and at anytime using the best technology that is physically possible to employ \footnote{All three criteria should be taken to their limits \textit{together} for an entity to fully pass the test for nonexistence. Take for example the ``anytime'' criteria.  Consider not the present, but the technology we have been able to develop/harness billions of years in the future, such that technology will truly be at the limits that the laws of physics allow.}.  I have found that this high bar for exclusion from reality is achieved by very few things, but importantly, we show in this section that it is clearly achieved for spatial extents less than $2l_p=3.24 \times 10^{-35}$ m and temporal durations less than $2\tau_p=1.08 \times 10^{-43}$ s, where $l_p=1.62 \times 10^{-35}$ m and $\tau_p=5.39\times 10^{-44}$ s are the Planck length and Planck time respectively.  What follows is the derivation of this smallest measurable length and this shortest measurable temporal duration and their subsequent associations with the atom of space and the atom of time respectively \footnote{From a LP perspective, proving that there is a minimal spatial extent and a minimal temporal duration implies discreteness of space and time respectively.  Outside of LP, these connections of minimal extent and duration to discreteness of ST are problematic \citep[438]{Kragh1994}.}.  First however, a note is in order on the use of NA-space.  

NA-space is required because it solves the anisotropy problem normally encountered with (or inherent to) other DST models that include \textit{a priori} existence of a grid.  However, the concept of NA-space is not new to the study of DST.  Rather than calling it NA-space, Silberstein avoids those words and instead calls it a mere ``labeling'', as recounted by Kragh \citep[459]{Kragh1994} \footnote{We agree with Silberstein and view space and time coordinates as simple parameters in the wavefunction describing a particle. However, one should be careful to note that distances in \textit{any} direction come in steps of $\chi$, not just distances along arbitrarily chosen \textit{x}, \textit{y} and \textit{z} directions.}:    

\begin{quotation}
	Time and space were viewed merely as a system of labeling [sic] events by numbers (x,y,z;t). If these labels were restricted to integers, space-time was said to be discrete. This is a more radical, but also a logically more satisfactory view than the one held by time-atomists such as Pokrowski. In particular, it avoided the problem of how to define durations and extensions without making use of continuous space-time as a background reference (\cite{Schild1949}).
\end{quotation}

\begin{figure}[H]
	\centering\includegraphics[width=10cm]{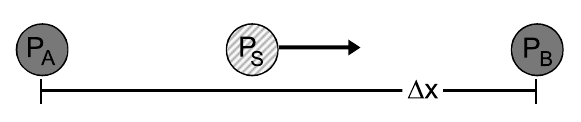}
	\caption{To measure distances, the most accurate ``ruler'' consists of two probe-particles $P_A$ and $P_B$ (that are stationary relative to each other) between which the spatial interval $\Delta x$ is to be measured.  $P_A$ emits the signal particle $P_S$, and $P_B$ receives $P_S$.}
	\label{fig:Fig6}
\end{figure}

\begin{figure}[H]
	\centering\includegraphics[width=10cm]{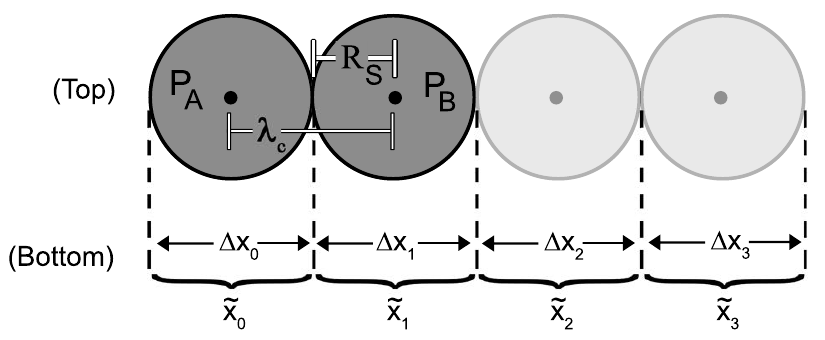}
	\caption{\textbf{Top:} A system to measure the smallest spatial separation between two distinct probe-particles (in darker gray). Note that $P_S$ is not shown in this figure.  The probe-particles need to remain entirely distinct, as spatially compact as possible, and be able to emit a signal particle $P_S$.  \textbf{Bottom:} The set of all continuous space x-values (denoted as $\Delta x_n$) within each sphere that are mapped to single x-values in discrete space, namely $\tilde{x}_n$.}
	\label{fig:Fig7}
\end{figure}

Now back to the heuristic derivation of $\chi$ and $\tau$.  One first considers how distances are measured.  Einstein instructed us that the best way to measure a distance is with light, by having one probe (P\textsubscript{A}) emit a signal photon (P\textsubscript{S}) and another probe (P\textsubscript{B}) receive P\textsubscript{S}, as shown in Fig. \ref{fig:Fig6}.  The distance between P\textsubscript{A} and P\textsubscript{B} is then equal to the transit time divided by the speed of light $c$.  One then constructs the probes such that they measure the shortest possible distance.  Besides placing the probes as close to each other as possible (Fig. \ref{fig:Fig7}), you would also strive to make P\textsubscript{A} and P\textsubscript{B} as small as possible. This is because one does not know from where within P\textsubscript{A} the probe signal P\textsubscript{S} is created or where P\textsubscript{S} is received within P\textsubscript{B}. According to quantum mechanics, with its concept of a particle's Compton wavelength ($\lambda_c=\hbar /mc$) being the limit on the measurement of the position of a particle, to make P\textsubscript{A} and P\textsubscript{B} as small as possible, you would choose very massive particles to serve as these probes.  But if they are too massive, they will be black holes from which P\textsubscript{S} cannot escape and perform the measurement.  The balance between localizing the probes P\textsubscript{A} and P\textsubscript{B} while still allowing P\textsubscript{S} to travel from one probe to the other is struck when the diameters of the probes ($D_{min}$) equal their Compton wavelengths \textit{and} equal twice the Schwartzchild radius $R_s=2Gm/c^2$.  Thus, setting $\lambda_c=2R_S$ yields a miminum diameter for P\textsubscript{A} and P\textsubscript{B} of:

\begin{equation}\label{chronon}
D_{min}=2\sqrt{\hbar G / c^3}=2l_p
\end{equation}

\hspace{0.75cm} \rule{9.15cm}{1pt}
\vspace{0.75cm}

As stated before, to measure the shortest possible length, one places the two probes P\textsubscript{A} and P\textsubscript{B} as close as possible to each other while ensuring that they remain distinct (Fig. \ref{fig:Fig7}).  The center-to-center separation of the two probes is then the smallest measurable length, namely, the hodon $\chi$. With $\chi$, one can calculate the mass of the probe $m_o$.  Both $\chi$ and $m_o$ are given below.   

\begin{subequations}
\begin{align}
&\chi = 2 \sqrt{\frac{\hbar G}{c^3}}=2 l_p = 3.23 \times 10^{-35} m \label{choron} \\
&m_o = \frac{1}{2}\sqrt{\frac{\hbar c}{G}}=\frac{1}{2}m_p= 1.09 \times 10^{-8} kg \label{mass_probe}
\end{align}
\end{subequations}

Now that $\chi$ has been determined, there are a couple of approaches that can be used to determine the shortest measurable temporal duration $\tau$.  One approach is to strictly adhere to LP and state that since one does not know from ``where'' within each spatial atom the photon is emitted, the resolution of temporal measurements is at best $\tau=\chi / c=2\tau_p$.  Another approach involves arguing that since we have not observed any particle traveling faster than the speed of light, the speed of light is indeed the maximum possible velocity \footnote{If a particle is found to travel faster than light, then its velocity would replace $c$ in the Lorentz factor. This is because this speedier particle would replace the photon in the ideal clocks used to calculate the Lorentz factors.  However, time dilations (for durations much larger than $\tau_p$) have been experimentally verified and are in agreement with a Lorentz factor that has $c$ for the maximum velocity. Thus, we can safely assume $c$ is the maximum allowable velocity.}.  And because the fastest travel possible in DST is a particle translating one $\chi$ for every duration $\tau$, the chronon is simply $\tau=\chi / c$.  Both approaches yield:

\begin{equation}\label{chronon}
\tau =\frac{2 l_p}{c}=2\tau_p=1.08 \times 10^{-43} s,
\end{equation}

\hspace{0.75cm} \rule{9.15cm}{1pt}
\vspace{0.25cm}

If one is unhappy with this LP-based derivation, one can consider other derivations given by (\citet{Sorkin1983, Riggs2009, Ng1995, Misner1973}) that yield similar values.  Even though I believe in the restrictive form of LP described in this paper, I view the calculation developed in this section as a simple heuristic argument and derivation of $\chi$ and $\tau$.  I provide it more as a salve to those married to the conventional view of space and time; it is preferred to view $\chi$ and $\tau$ not as derived quantities, but rather as \textit{constants of nature} \footnote{At least viewing $\chi$ and $\tau$ as constants when gravity is absent.  As we talk about in Section \ref{sec:Discussion}, perhaps gravity can affect the values of $\chi$ and $\tau$. This would make the atomic picture of space being more granular than atomistic, with each grain potentially being of different size depending on the amount of mass or gravitational potential energy within the grain.}, or within loop quantum gravity, as the minimal eigenvalues of quantum observables (\citet{Rovelli2003}).

\subsubsection{Leopold's Theorem}
\begin{figure}
	\centering\includegraphics[width=4cm]{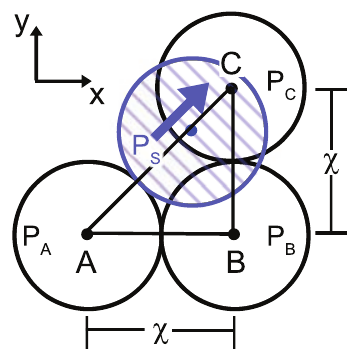}
	\caption{Distances are measured along each path according to the rules described in the text. For this particular triangle with $\overline{AB}=\overline{BC}=\chi$, $\overline{AC}$ equals $\chi$ as well.  This is because only one jump of $\chi$ is needed along the diagonal for the sphere defining $P_S$ to partially overlap the sphere centered about $C$, and therefore be at the same position in discrete space.}
	\label{fig:Fig8}
\end{figure}

With the value of $\chi$ now in hand, we can start our approach in developing the DST distance formula.  To start, we reject the first step in Weyl's construction where he assumed absolute space and drew a grid -- we instead assume NA-space. In NA-space, a particle can travel in \textit{any} direction as long as the magnitude of any individual translation is $\chi$.  We then construct a system to measure the distances of a triangle's sides and hypotenuse.  The system is composed of three particles $P_A$, $P_B$, $P_C$ at positions $A$, $B$, and $C$ respectively, with $P_A$, $P_B$ and $P_C$ able to emit or receive a signal-particle $P_S$ (Fig. \ref{fig:Fig8}). The particles $P_A$, $P_B$, $P_C$, and $P_S$ all have diameters equal to $\chi$.  

We first construct the smaller right triangle shown in Fig. \ref{fig:Fig8} such that the duration between emission (at $A$) and reception (at $B$) of $P_S$ is $\tau$ \footnote{Note that $P_S$ is only shown along the segment $A \rightarrow C$, but signal-particles also traverse the segments $A \rightarrow B$ and $B \rightarrow C$ when the measurement of these segments are performed.}.  This duration corresponds to a length for the path $\overline{AB}$ of $\chi$.  Additionally, the system is constructed such that a similar measurement yields $\chi$ as the length of the path $\overline{BC}$.  Thus the system is an isosceles right triangle with $\overline{AB}=\overline{BC}=\chi$.  However the length of the hypotenuse $\overline{AC}$ is \textit{not} $\sqrt{2} \chi$.  To see this, consider a signal-particle $P_S$ emitted by $P_A$ (centered about \textit{A}) towards $P_C$ (centered about \textit{C}).  $P_S$ makes its first discrete jump of $\chi$ and \textit{already} the sphere that specifies the position of $P_S$ partially overlaps with the sphere defining the position $C$.  Hence, $P_S$ has \textit{arrived} at $C$ and has been received by $P_C$.  This process takes the same duration $\tau$ as that required by $P_S$ traveling along the path $P_A$ $\rightarrow$ $P_B$.  Therefore, the length of the hypotenuse is equal to the lengths of the sides, all being $\chi$, and thus the Pythagorean theorem is violated.

\begin{figure}[H]
    \centering\includegraphics[width=6cm]{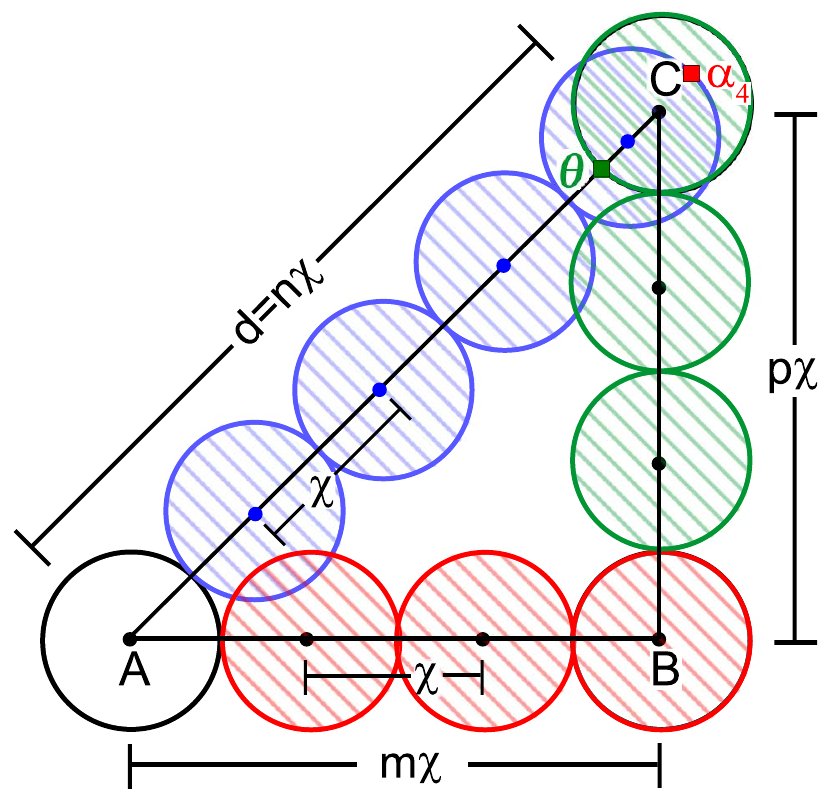}
    \caption{For an arbitrarily large triangle, the distance formula Eq. \eqref{LeopoldB} is easily derived by determining how many translations are required along $\overline{AC}$ such that the leading edge of the translated point along the hypotenuse (denoted by $\alpha_n$) overtakes the trailing edge (denoted by $\theta$) of the sphere that defines point $C$.}
    \label{fig:Fig9}
\end{figure}

Next, we consider an arbitrarily large right triangle with the lengths of the two sides as $m\chi$ and $p\chi$, as shown in Fig. \ref{fig:Fig9}.  It is easy to derive an equation for the distances from $A$ to $\alpha_n$ and from $A$ to $\theta$, where $n$ is the jump number, $\alpha_n$ is the leading edge of the translating point, and $\theta$ is the trailing edge (i.e., closest to $A$) of point $C$ in Fig. \ref{fig:Fig9}. Then using these equations, one can determine the smallest number of jumps ($n$) necessary for $P_S$ to arrive at point $C$:

\begin{align}
	n &> \sqrt{m^2+p^2} - 1 \label{LeopoldB} \\
	  &= \left \lfloor \sqrt{m^2+p^2} \right \rfloor \nonumber
\end{align}

\noindent where $m$ and $p$ are integers, $m\chi$ and $p\chi$ are the lengths of base and height respectively, and $n$ is the smallest integer that satisfies Eq. \eqref{LeopoldB}.  It is seen that for triangles with $m=p=1$ and $m=p=2$, the lengths of the hypotenuses are equal to the lengths of the sides.  However, as the sides of a right triangle become larger (as $m$ and $p$ become large), the length of the hypotenuse converges to $\sqrt{2}$ times the length of the side and the Pythagorean theorem is restored. Also, note that Eq. \eqref{LeopoldB} is identical to Eq. \eqref{BendegemGeneralC}.  How this equation conserves the immutability of the atoms of space and time is discussed in Section \ref{sec:TimeLength}.

Already, we see that this model contains several attractive properties:

\begin{enumerate}
	\item\label{Advantage 1} It fully embraces the concept of NA-space, thereby maintaining isotropy.
	\item\label{Advantage 2} Measurements of lengths are performed in ways accepted by science and adheres to the tenets of a conservative form of LP.
	\item\label{Advantage 3} A single equation applies to all size-scales and satisfies both the Pythagorean theorem for any practical distance and the requirement of discretized space (i.e., distances being integer multiples of $\chi$).
\end{enumerate}

The three-dimensional version of the new distance formula (called Leopold's theorem) in units of distance is

\begin{align}
n &> \sqrt{m^2+p^2+s^2} - 1 \label{Leopold} \\
n &= \left \lfloor \sqrt{m^2+p^2+s^2} \right \rfloor \nonumber
\end{align}

\noindent where $m$, $p$ and $s$ are integers, $m\chi$, $p\chi$ and $s\chi$ are the lengths of the three constituent segments, and $n$ is the smallest integer that satisfies Eq. \eqref{Leopold}.

\subsection{Continuous-Discrete Associations or Mappings}
Now that the atoms of space and time are derived, we can express the associations of spatial positions in continuous space-time (CST) to values in discrete space-time (DST), along with the CST/DST associations for time:

\begin{subequations}
\begin{align}
\tilde{x}_m &= \left ( \left(m-\frac{1}{2} \right ) \chi, \left( m+\frac{1}{2} \right ) \chi \right ]\label{space_mapping} \\
\tilde{t}_n &= \left [ n \tau, \left(n+1 \right) \tau \right )\label{time_mapping} 
\end{align}
\end{subequations}

\noindent with $m$ being any positive or negative integer and $n$ being an integer greater than or equal to zero. Thus, $\tilde{x}$ and $\tilde{t}$ are always intervals, or continuous sets of CST counterparts.  But one must remember that the intervals defined by Eq. \eqref{space_mapping} are along a path and not along a particular direction or arbitrarily chosen axis.  \footnote{It is also important to note that there are no ``gaps'' in space, i.e., volumes that are unoccupied by atoms of space.  Upon remembering that this model assumes non-absolute space, and with the proper implementation of the model, one sees that all points in continuous space can be assigned DST coordinates.}

Describing how velocity is affected in DST is important for later sections of this paper on special relativity.  However, a comprehensive discussion of velocity is a bit lengthy and not essential for the main subjects of this paper, it is therefore relegated to the appendix.  The important thing to remember is that CST values for velocities are artificial--it is the DST concept of velocity that is real.  Please see the appendix for details.

\subsection{Comparison of the Three Distance Formulas}
It is seen that all three approaches, while using different starting points and assumptions, lead to the same equation for the calculation of distances in discrete space (Eqs. \eqref{BendegemGeneralC} or \eqref{LeopoldB}). After assigning $N_D$ to be equal to one, and allowing the grid to change orientation dependent on the direction of a particle's travel, Van Bendegem's approach matches that of Crouse's.  Also, it is seen that within Forrest's approach are aspects of Van Bendegem's and Crouse's approaches.  Forrest's identification of points $\left \langle u,v \right \rangle$ creates a \textit{de facto} and \textit{a priori} existing grid similar to, but playing a lesser role than the grid in Van Bendegem's approach.  Finally, Forrest's BoAs are similar to the ``atoms of space'' used by Crouse.  Contrary to Crouse's approach however, within Forrest's BoAs are a multitude of identifiable and unique grid points - something we believe is entirely inconsistent with the concept of an atom of space.  Even if the grid is only a mathematical tool, the use of it clouds important aspects of the true nature of DST (e.g., its inherent isotropy and non absolute nature). Before using Eq. \eqref{LeopoldB} in the new derivation of time dilation and length contraction, we briefly discuss an interesting mathematical property of the new distance formula.

\subsection{\label{subsec:Triangle}Violation of the Triangle Inequality Theorem}
The triangle inequality (TI) theorem is one of the most sacrosanct theorems in all of mathematics.  The inviolable nature of the TI theorem is inculcated into everyone, from beginning students of mathematics: ``the single most important inequality in analysis is the triangle inequality'' (\citet{Kaye2015}), to seasoned practitioners of the craft: Forrest in pure mathematics \citep[329]{Forrest1995} and Brightwell in causal set theory (\citet{Brightwell1991}) to name only two of many. The TI theorem states that the distances between points $e$, $f$ and $g$, namely $d(e,f)$, $d(f,g)$ and $d(e,g)$, satisfy the inequality $d(e,g) \leq d(e,f)+d(f,g)$. Forrest was no different than everyone else in testing his model's adherence to the TI theorem, and being reassured by its passing. Forrest achieves compliance with the theorem because the distances his method yield are generally larger than those predicted by Eq. \eqref{LeopoldB}.  However, we will see in this section and the next that all metrics that adhere to the TI theorem yield non-physical distances in DST and cannot be used to accurately calculate distances in the real space in which we live (or at least distances at the Planck scale).   

The non-adherence of Eq. \eqref{LeopoldB} to the TI theorem may be somewhat unexpected, but it is necessary to conserve the the immutability of atoms of space and time while also having distances converge to values predicted by the Pythagorean theorem at macro-geometic scales.  To see this violation of the TI theorem, consider three collinear link-points $e=[0,0]$, $f=[1,1]$ and $g=[3,3]$ in Fig. \ref{fig:Fig5} (see Section \ref{sec:Forrest} for a description of the labeling system). Upon studying Fig. \ref{fig:Fig5}, one sees that Forrest's \textit{original} approach yields $d(e,f)=2$, $d(f,g)=3$ and $d(e,g)=5$, thus satisfying the TI theorem.  However, Van Bendegem's, Crouse's, and Forrest's modified approaches (all resulting in Eq. \eqref{LeopoldB}) yield $d(e,f)=1$, $d(f,g)=2$ and $d(e,g)=4$ which does not satisfy the TI theorem \footnote{This only occurs for segments off of the arbitrarily chosen \textit{x}, \textit{y}, \textit{z} axes; the metric defined by Eq. \eqref{LeopoldB} always gives $d(e,g) = d(e,f)+d(f,g)$ for points \textit{e}, \textit{f} and \textit{g} all being on one of the three axes.}.  One can speculate that Forrest was pleased with his result (namely, adherence to the TI theorem for the smallest possible distances) because in \citep[329]{Forrest1995} he states ``to be sure, we would not call a quantitative relation \textit{distance} unless it satisfied the triangle inequality''.  However, Forrest's result of $d(e,f)=2$ should have raised red flags; it will be shown in the next section that such a result does not conserve the immutable nature of the atom of space, in contradistinction to Eq. \eqref{LeopoldB}.  In fact, upon further study, one realizes that the only models of DST that conserve the atoms of space and time \textit{are ones that violate the TI theorem}.  Namely, in flat-space, one will always encounter cases (especially at the Planck scale) where a smaller value is obtained for the sum of the distances of component segments relative to the value of the distance of the parent segment \footnote{Again, we are only considering flat-space in this paper.}.

\section{\label{sec:TimeLength}Time Dilation and Length Contraction in Discrete Space-Time}
\begin{figure}[H]
	\centering\includegraphics[width=0.70\textwidth]{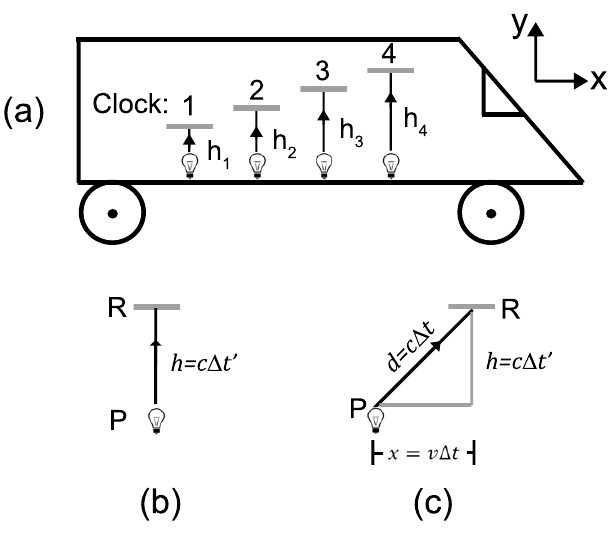}
	\caption{\textbf{(a):} An array of light clocks on a train traveling at a velocity $v$.  The clocks have values of $h$ as integer multiples of $\chi$.  \textbf{(b):} One of the clocks from the perspective of an observer in RF2.  \textbf{(c):} One of the clocks from the perspective of an observer in RF1.}.  
	\label{fig:Fig10}
\end{figure}

Consider the standard derivation for time dilation given in any textbook on SR \citep[chap. 4]{Helliwell2010} that involves two observers and an ideal ``light-clock'' on a train traveling in the $\hat{x}$ direction, as shown in Fig. \ref{fig:Fig10}.  We update this calculation slightly by replacing the mirror that is typically used in the clock with a photon receiver (\textbf{R}). Similar to the mirror in the standard derivation, \textbf{R} is placed above the photon emitter (\textbf{P}) in Fig. \ref{fig:Fig10}.  This change allows us to assess shorter time durations.  Also, instead of one light-clock with a height $h$, we consider a collection of light-clocks with $h=n'\chi$ with $n' \in \{1,2,\dots \}$, as shown in Fig. \ref{fig:Fig10}.  All the clocks are at $x=x'=0$ at $t=t'=0$, and are identified from here on according to their value of $n'$.  Unprimed values $\Delta t$ and $\Delta x$ correspond to temporal durations and spatial extents in reference frame 1 (RF1) as recorded by an observer (O1) alongside and stationary relative to the train tracks. Primed values are used for reference frame 2 (RF2), namely, values that are recorded by an observer (O2) on the train.

\begin{figure}[H]
	\centering\includegraphics[width=10cm]{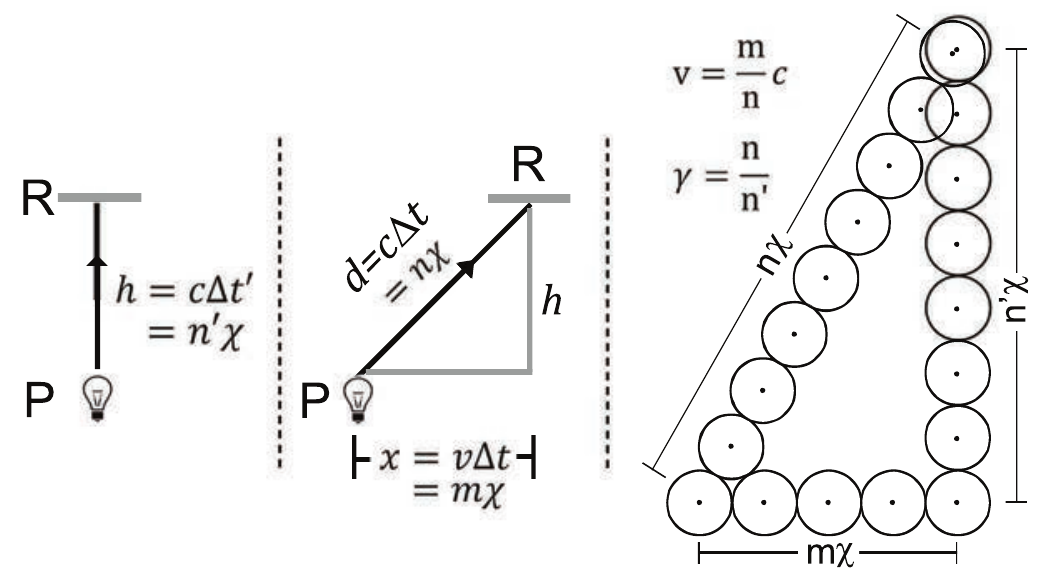}
	\caption{\textbf{Left:} One of the clocks from the perspective of an observer in RF2.  \textbf{Center:} The same clock from the perspective of an observer in RF1.  \textbf{Right:} The DST representation of the sides of the triangle defined by the path of the photon (in both RFs) and the motion of the clock.}
	\label{fig:Fig11}
\end{figure}

Now consider the trajectory of a photon from O1's and O2's perspectives, remembering the fact that a photon travels at a velocity $c$ in both RFs.  For any clock $n'$, the time elapsed while a photon travels from the emitter to the receiver is $\Delta t'=n'\chi /c$ in RF2 and $\Delta t = n\chi /c$ in RF1, during which time the clock has moved $\Delta x = m\chi =v \Delta t$ where $v$ is the velocity of the train (again, see Fig. \ref{fig:Fig10} and \ref{fig:Fig11}).  Also, and as typical, $\Delta y = \Delta y' = n' \chi$.  Thus, we have the standard right triangle with the lengths of the sides as $m\chi$ and $n'\chi$ and the length of the hypotenuse as $n\chi$, with $v$ and $\gamma = \Delta t / \Delta t'$ given by

\begin{subequations}
\begin{align}
v &= \frac{m}{n}c \hspace{10mm} \\
\gamma &= \frac{n}{n'}\label{gamma}.
\end{align}
\end{subequations}

At this point in the \textit{conventional} calculation, one would use the distance formula given by the Pythagorean theorem:

\begin{align}	
n^2 \chi^2 &= m^2 \chi^2 + n'^2 \chi^2 \hspace{30pt} \nonumber  \\
\rightarrow  \hspace{30pt}      n^2  &= m^2  + n'^2  \label{PythagorasB}
\end{align}

After using the relations $m \chi = v \Delta t$, $n \chi =c \Delta t$, $n' \chi =c \Delta t'$, we can easily solve for the relation between $\Delta t$ and $\Delta t'$ as first derived by Einstein, Poincar\'e and Lorentz:

\begin{subequations}
	\begin{align}
\Delta t &= \frac{1}{\sqrt{1-v^2/c^2}} \Delta t' = \gamma_E \Delta t' \label{time_einsteinA} \\
n &= \gamma_E \hspace{2pt} n' \label{time_einsteinB}
\end{align}
\end{subequations}

\noindent where the subscript $E$ stands for Einstein. Length contraction would then be typically given as: 

\begin{subequations}
\begin{align}
\Delta l &= \gamma_E \Delta l'  \label{length_einsteinA} \\
m &= \gamma_E \hspace{2pt} m'    \label{length_einsteinB}
\end{align}
\end{subequations}

\noindent where $\Delta l=m\chi$ is the length of a rod (or a distance), as measured by an observer stationary relative to the rod, i.e., the ``proper length" of the rod. The term $\Delta l'=m'\chi$ is the observed length of the rod, namely, the length measured by an observer traveling at a velocity $v$ relative to the rod.

Everything done so far in the calculation leading to Eqs. \eqref{time_einsteinA}-\eqref{length_einsteinB} is currently accepted by the scientific community almost without question \footnote{This was not the case prior to the 1922 Einstein-Bergson debate, when many philosophers and physicists viewed Einstein's theory and interpretation with a high degree of skepticism (\citet{Canales2015}).}.  However, Eqs \eqref{time_einsteinA}-\eqref{length_einsteinB} are the roots of the oft-cited problem concerning the variability of $\chi$ and $\tau$ \citep[69-71]{Hagar2014} \footnote{A typical example of the lack of serious thought and debate on the subject of DST by recognized authorities on quantum gravity is well exhibited in (\cite{Hythloday2017}).}. This is because people generally assume that $v$ can be \textit{any number} between $0$ and $c$, and a \textit{constant number} at that, namely, independent of time (i.e., independent of $n$ and $n'$). However, $v=mc/n$, with $m$ and $n$ being integers. Even though $m$ changes as $n$ increases so as to keep $v$ as close as possible to some desired value, the system will never be able to maintain a constant value for $v$ as $n$ increases (unless of course $m=0$ or $m=n$ for all $n$). Building upon this faulty but commonly held assumption of $v$ being independent of time, one would then conclude that $\gamma_E$ is independent of time.  And if this is the case, then all temporal durations $\Delta t'$ measured by all the clocks on the train are dilated by this same factor $\gamma_E$ in RF1.  This all leads to the incorrect belief that the atom of time $\tau$ in RF2 is dilated to a larger value in RF1 and the atom of space $\chi$ in RF1 is contracted to smaller value in RF2. This perceived contradiction has stymied DST's serious study for over 100 years.  Additionally, without a formal procedure to calculate distances in DST, how do you handle Eq. \eqref{PythagorasB}?  The terms $n$, $m$ and $n'$ all need to be integers, thus one is tempted to consider either a floor or ceiling operation on Eq. \eqref{PythagorasB} to achieve integer values:

\begin{subequations}
	\begin{align}
	n &= \floor{\sqrt{m^2+n'^2}} \label{PythagorasFixA} \\
	n &= \ceil{\sqrt{m^2+n'^2}} \label{PythagorasFixB}
	\end{align}
\end{subequations}

\noindent Of these two equations, most people would naturally prefer Eq. \eqref{PythagorasFixB} because the distances it calculates always satisfy the triangle inequality, as opposed to Eq. \eqref{PythagorasFixA}.  However, Eq. \eqref{PythagorasFixB} yield the same results as Forrest's method and lead to one important nonphysical result discussed later in this work; Eq. \eqref{PythagorasFixA} is identical to Eqs. \eqref{BendegemGeneralC} and \eqref{LeopoldB} and we argue in the rest of this work that it is the correct choice.  But rather than guess which equation to use (Eq. \eqref{PythagorasFixA} or Eq. \eqref{PythagorasFixB}), one should use the formal approach that led to Leopold's theorem (Eq. \eqref{LeopoldB}) and use the concepts of DST and non-absolute space discussed earlier and throughout this work.  Upon doing so, one sees that the problems with the velocity dependent extent (duration) of the atom of space (time) are solved.

Our new derivation of time dilation in DST starts the same way as the conventional calculation, with light-clocks on the train.  However, in the new derivation, each light-clock of different height $n'\chi$ assesses a different temporal duration $\Delta t' = n'\chi / c$, starting from $\Delta t'=\chi /c = \tau$ (the shortest possible duration) to progressively larger integer multiples of $\tau$.  The only other change in the derivation is the use of Eq. \eqref{LeopoldB} instead of the Pythagorean theorem.  To start, you decide the velocity $v$ for which you will calculate $\gamma$; let us consider $v=0.5c$ as an example (note that $\gamma_E = 1.15$ for this velocity).  Note that for particular durations it is not always possible to have $v$ be exactly equal to this, or some other desired value.  In these cases, we adopt a convention (called the assignment rule later in this work) where we choose the value of $m$ such that $mc/n$ is as close to but smaller than (or equal to) the desired value, namely, $m=\floor{(v/c)n}$.  But this convention is not necessary, and one is free to choose a different convention.  Alternatively, in DST, a velocity value $v$ (or rather $v/c$) should be viewed as a probability that a particle makes a spatial translation of $\chi$ over the duration of a tick $\tau$ (see the appendix for details on velocity in DST). The rest of the procedure is described next.

Once the velocity is chosen, one next sets all the clocks to $t=t'=0$, at which time all their emitters \textbf{P} emit a single photon towards their receivers \textbf{R}.  Then one assesses the situation when the receiver for clock $n'=1$ detects the photon at $\Delta t'=\tau$.  From the perspective of $O1$, the clock could either have made a spatial translation of extent $\Delta x = \chi$ or not moved at all - these are the only two possibilities.  If it did move by $\chi$, then this corresponds to a velocity of $v=c$ over this duration; if it did not move, the velocity is assessed to be zero.  To be consistent with our convention of $v$ being less than or equal to $0.5c$, we choose the $v=0$ case.  Solving Eq. \eqref{LeopoldB} yields $n=1$; and with $\gamma = n/n'$, we find that $\gamma = 1$ for this duration.  Even if we had chosen the case where $v=c$, Eq. \eqref{LeopoldB} yields $n=1$.  Therefore $\gamma=1$ for a duration of $\Delta t'=\tau$ \textit{regardless} of the relative velocity of the two RFs (again, $v=0$ and $v=c$ are the only two possible velocities for a time duration of $\tau$).

\begin{figure}[H]
		\centering\includegraphics[width=11cm]{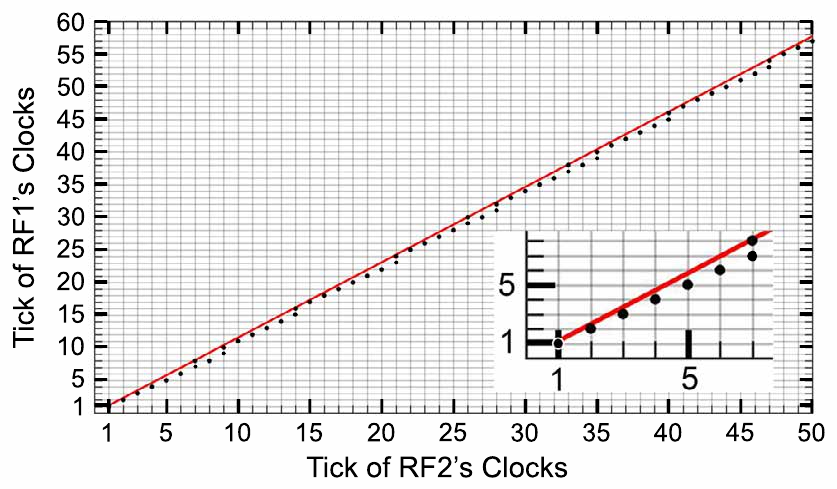}
		\caption{The correspondence between the ticks of the clocks in RF1 to the ticks of the clocks in RF2. The RFs have a relative velocity of $0.5c$.  The red line shows $\gamma_E=1.15$.  \textbf{Inset}: No time dilation occurs for the first seven ticks (each of duration $\tau$), but then the clocks on the train start trailing the clocks at the station.}
		\label{fig:Fig12}
	\end{figure}

\begin{figure}[H]
	\centering\includegraphics[width=1\textwidth]{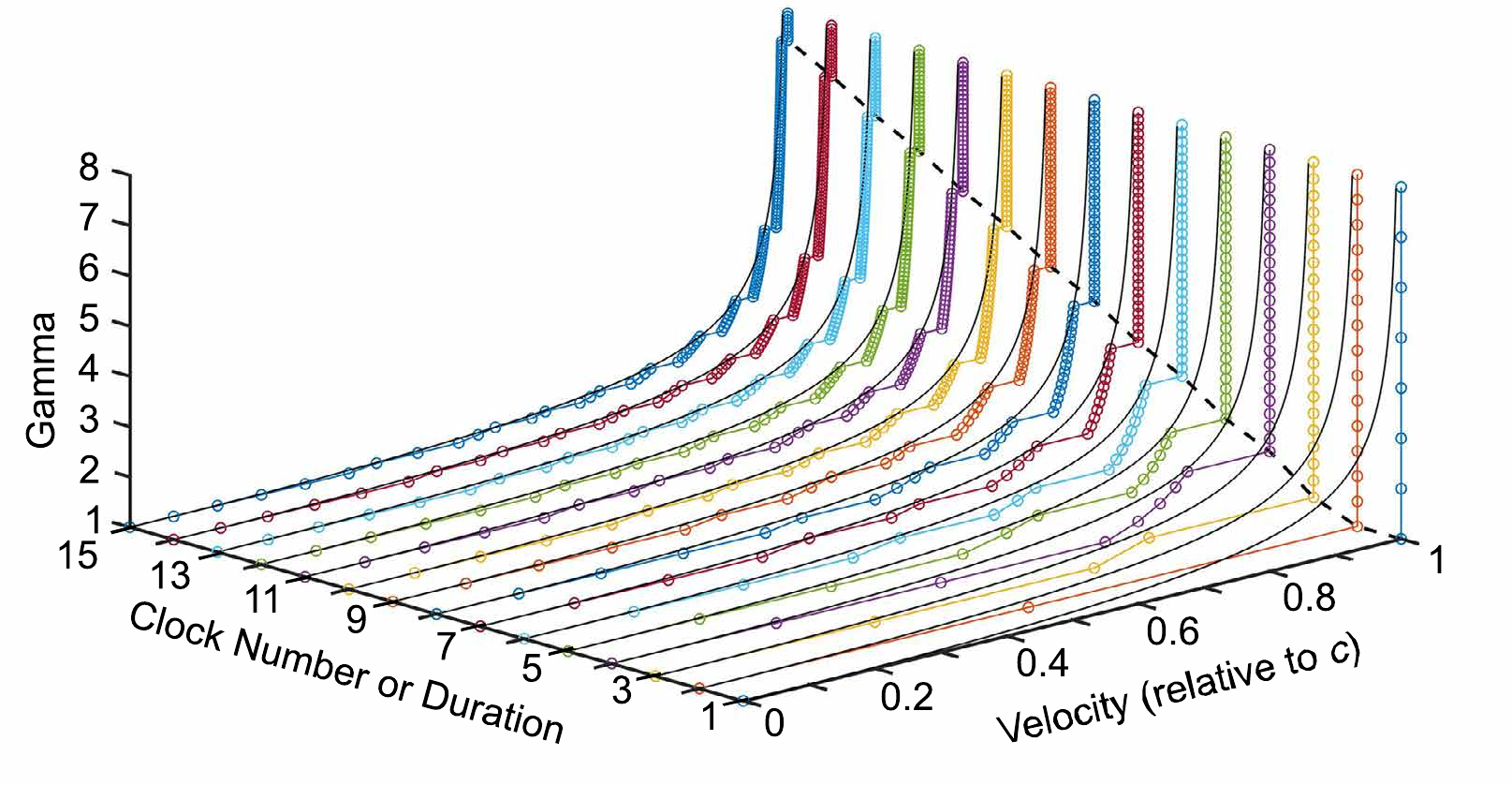}
	\caption{The values of $\gamma$ versus duration ($n'$) and velocity (colored solid lines) with allowable values indicated by the markers ``$\circ$'', and $\gamma_{critical}$ shown with the dotted black line.  Also shown is $\gamma_E$ versus continuous values of $v$ (solid black).  Velocity is given relative to $c$, and duration is relative to $\tau$. Note that a velocity equal to $c$ is allowed for any duration, even for a particle with nonzero rest mass. Also note that while $c$ is the maximum velocity, $\gamma$ can increase without limit as ever more energy is provided to the system.}
	\label{fig:Fig13}
\end{figure}

We next assess the situation at $\Delta t'=2\tau$ when the receiver of clock $n'=2$ detects its photon.  A solution exists for Eq. \eqref{LeopoldB} with $n'=2$, $m=1$ and $n=2$, corresponding to a velocity of $v=0.5c$ and $\gamma=1$.  Thus, a duration of $\Delta t'=2\tau$ in RF2 corresponds to the same duration $\Delta t=2\tau$ in RF1.  For clock $n'=3$, no solution to Eq. \eqref{LeopoldB} exists that has $v=0.5c$. Two solutions with $v$ closest to $v=0.5c$ are: $\{ n',m,n \} = \{ 3,1,3 \}$ with $v=0.33c$, and $\{ n',m,n \} = \{ 3,2,3 \}$ with $v=0.67c$.  We choose the $\{ 3,1,3 \}$ solution.  We finish this procedure by recording the results from the first 30 clocks in Table \ref{table:Table1} and the first 50 clocks in Fig. \ref{fig:Fig12}. Figure \ref{fig:Fig13} shows $\gamma$ for clock durations of $\Delta t'=1\tau \rightarrow 15\tau$ for all possible velocities, including $v=c$ (note that travel at the speed of light is possible in DST, as discussed in Section \ref{sec:LightSpeed}).

One encounters a curious situation in DST that has no continuous space-time equivalent: lack of a one-to-one correspondence of the ticks of RF1's clock to those of RF2's clock. In continuous space-time, one can always imagine clocks in RF1 with the necessary tick rate to have a one-to-one tick correspondence with the ticks of clocks in RF2.  In DST however, the highest tick rate allowed is one tick per $\tau$ of duration.  Thus, it is not surprising to find instances where two ticks of RF1's clocks occur for one tick of RF2's clock, for example, the $7^{th}$, $9^{th}$ and $14^{th}$ tick of RF2's clock.  What is surprising though is that even though in the long-term RF1's clocks tick rate is faster, there are instances where two ticks of RF2's clock can transpire for a single tick of RF1's clock. This occurs for the $8^{th}$ tick of RF1's clock.  Thus, DST takes even further the inherent asynchronous nature of clocks in conventional SR; not only can you not synchronize a set of three clocks (two of which are moving and one is stationary), you cannot even find a one-to-one correspondence between two clocks (moving relative to each other). 

\begin{table}[H]
	\caption{The height, base and hypotenuse (relative to $\chi$) of the triangles traced out by the photons in the light-clocks, the velocity (relative to $c$), and $\gamma$. The correspondence between the ticks of the clocks in RF1 and RF2 is also given.}

	\begin{tabular}{|c|c|c|c|c|c|}
		\hline
		Height ($n'$) or& Base& Hypotenuse ($n$) or& $v$& $\gamma (v,n')$& Contracted\\
		tick of RF2's clock& ($m$) & tick of RF1's clock& & &length (m')\\
		\hline
		1 & 0 & 1 & 0 & 1 & 0 \\
		\hline
		2 & 1 & 2 & 0.5 & 1 & 1 \\
		\hline
		3 & 1 & 3 & 0.33 & 1 & 1 \\
		\hline
		4 & 2 & 4 & 0.5 & 1 & 2 \\
		\hline
		5 & 2 & 5 & 0.4 & 1 & 2 \\
		\hline
		6 & 3 & 6 & 0.5 & 1 & 3 \\
		\hline
		7 & 3 & 7 & 0.43 & 1 & 3 \\
		\hline
		7 & 4 & 8 & 0.5 & 1.14 & 3 \\
		\hline
		8 & 4 & 8 & 0.5 & 1 & 4 \\
		\hline
		9 & 4 & 9 & 0.44 & 1 & 4 \\
		\hline
		9 & 5 & 10 & 0.5 & 1.11 & 4 \\
		\hline
		10 & 5 & 11 & 0.45 & 1.10 & 5 \\
		\hline
		11 & 6 & 12 & 0.5 & 1.09 & 5 \\
		\hline
		12 & 6 & 13 & 0.46 & 1.08 & 6 \\
		\hline
		13 & 7 & 14 & 0.5 & 1.08 & 6 \\
		\hline
		14 & 7 & 15 & 0.47 & 1.07 & 7 \\
		\hline
		14 & 8 & 16 & 0.5 & 1.14 & 7\\
		\hline
		15 & 8 & 17 & 0.47 & 1.13 & 7 \\
		\hline
		16 & 8 & 17 & 0.47 & 1.06 & 8 \\
		\hline	
		16 & 9 & 18 & 0.5 & 1.13 & 8 \\
		\hline
		17 & 9 & 19 & 0.47 & 1.12 & 8 \\
		\hline
		18 & 10 & 20 & 0.5 & 1.11 & 9 \\
		\hline
		19 & 10 & 21 & 0.48 & 1.11 & 9 \\
		\hline
		20 & 11 & 22 & 0.5 & 1.10 & 10 \\
		\hline
		21 & 11 & 23 & 0.48 & 1.10 & 10 \\
		\hline
		21 & 12 & 24 & 0.5 & 1.14 & 10 \\
		\hline
		22 & 12 & 25 & 0.48 & 1.14 & 11 \\
		\hline
		23 & 12 & 25 & 0.48 & 1.09 & 11 \\
		\hline
		23 & 13 & 26 & 0.5 & 1.13 & 11 \\
		\hline
		24 & 13 & 27 & 0.48 & 1.13 & 12 \\
		\hline
		25 & 14 & 28 & 0.5 & 1.12 & 12 \\
		\hline
		26 & 14 & 29 & 0.48 & 1.12 & 13 \\
		\hline
		27 & 15 & 30 & 0.5 & 1.11 & 13 \\
		\hline
		28 & 15 & 31 & 0.48 & 1.11 & 14 \\
		\hline
		28 & 16 & 32 & 0.5 & 1.14 & 14 \\
		\hline
		29 & 16 & 33 & 0.48 & 1.14 & 14 \\
		\hline
	\end{tabular}
	\label{table:Table1}
\end{table}
 
For the reader's convenience, the following list contains the steps needed to calculate $\gamma$ as a function of $v$ and $\Delta t'$ \footnote{A Matlab program to implement the procedure is available upon request.}:

\begin{enumerate}
	\item Decide what velocity you want to use in the calculation \footnote{Again, note that it will not always be possible to have this velocity, and you may have to choose between a velocity less than, or greater than the desired value.}. 
	\item Choose a particular value of $n'$, corresponding to the duration $\Delta t'=n'\tau$ that will be used in the calculation.
	\item Use this value of $n'$ (from Step 2) and an array of $m$ values (i.e., $m=0,1,2, \cdots$) to calculate an array of corresponding $n$ values using Eq. \eqref{LeopoldB}.
	\item  Choose the $m$ value (and its associated $n$ value from Step 3) that yields a velocity ($v=mc/n$) closest to the desired value (from Step 1). 
	\item Calculate the Lorentz factor using the equation $\gamma=n / n'$.
	\item For any instance where a $n$ value is skipped, you flip this method -- rather than setting $n'$ and then finding $m$ and $n$, one sets $n$ to the desired value and then finds the appropriate values for $m$ and $n'$. This happens for $n$ values of 10, 16, 24, 26 and 32 in Table \ref{table:Table1}. 	
\end{enumerate}

Two important results can be gleaned at this point.  First, $\gamma$ converges to $\gamma_E$ for temporal durations large relative to $\tau$. It is interesting though that this convergence is not monotonic.  Second, a temporal duration of $\tau$ in RF2 is not dilated in RF1, i.e., it has a value of $\tau$ in RF1.  This will be true regardless of the velocity of the clocks, namely, $\gamma(v,\tau)=1$ for the only two possible values for velocity for this duration: $v=0$ and $v=c$.  Thus, the immutability of the atom of time is conserved.  This is not the case if one uses the distance formula derived by Forrest (his original formulation shown in Fig. \ref{fig:Fig4}); doing so yields a $\gamma$ factor of $\gamma(c,\tau)=2$, and hence the atom of time is dilated by a factor of two.  Thus Forrest's original formulation leads to this very important non-physical result; the modified approach (Fig. \ref{fig:Fig5}) does not suffer this problem.  Interestingly, Weyl's distance formula (Eq. \eqref{Weyl}) does conserve the immutability of the atom of time, but it leads to demonstrably incorrect results for time dilation for any duration larger than a few $\tau$ (i.e., it predicts no time dilation, regardless of a system's velocity).

While slightly more complicated, the same basic argument used in conventional SR to connect length contraction with time dilation can be used for length contraction in DST.  To summarize the argument (within the framework of DST), one comes to the conclusion that systems need to obey the principle of relativity.  That is, in a two particle system (i.e., particles $A$ and $B$) with particle $A$ advancing in the $+\hat{x}$ direction at a velocity $v$ from the perspective of, and relative to, particle $B$, one can equally say that particle $A$ is receding (i.e., in the $-\hat{x}$ direction) at a velocity $-v$ relative to particle $B$.  Thus, to determine the contracted length, i.e., $m'\chi$ in DST, we use the same ``assignment rule'' (i.e., the convention) that was used to determine the ``proper'' length $m\chi$, namely having $m'$ as the largest integer that still satisfies the condition $m'c/n' \leq v$, namely $m'=\floor{(v/c)n'}$.  But $v$ in this equation is not $mc/n$, but the value used in the rule when determining the values of $m$ for particular values of $n$; the example shown in this work used $v=0.5c$.  Thus, once you determine the integers $n$, $m$ and $n'$ in the calculation you have everything you need to assign the value of $m'$: for a particular value of $n'$, one simply looks up in Table \ref{table:Table1} what $m$ was when $n=n'$.  For example, to get the $m'$ value for $n'=15$, you look up the row in Table \ref{table:Table1} with $n=15$ and find that $m=7$ for this value of $n$. One can then complete a row of values:  $n'=15$, $m'=7$, $n=17$ and $m=8$.  To express this process in an equation, you would use the relation $n'/n=\gamma^{-1}$ in the length contraction equation $m'=\floor{(v/c)n'}$, resulting in:

\begin{equation}\label{length_contraction}
m' = \floor{\gamma^{-1} \frac{v}{c} n}
\end{equation}

\noindent Again, $v$ in Eq. \eqref{length_contraction} is not $mc/n$ but the target value used in the assignment rule (the convention).  This aspect of Eq. \eqref{length_contraction} insures strict adherence to the relativity principle for inertial reference frames.  Only in the limit of large $n$ does $v \approx mc/n$, in which case Eq. \eqref{length_contraction} reduces to the known and verified continuous ST equation of $m'=\gamma^{-1}m$. Thus, time dilation and length contraction in DST are given by:

\begin{align}
\Delta t' &= \gamma^{-1} \Delta t \label{timelengthdiscretea}, \\
\Delta l' &=  \left \lfloor \gamma^{-1} \frac{v}{c} \left \lceil \frac{c}{v}\frac{\Delta l}{\chi} \right \rceil \right \rfloor \chi \label{timelengthdiscreteb}.
\end{align}

\noindent with $\Delta t = n\tau$, $\Delta t' = n'\tau$, $\Delta l=m\chi$ is the proper length, $\Delta l' = m'\chi$ is the observed length (i.e., the contracted length); $\gamma = n/n'$, where $n$, $n'$, $m$ and $m'$ all integers, and $v$ is the desired constant velocity (e.g., $0.5c$).  We have also used the relation $n=\lceil cm/v \rceil = \lceil c\Delta l /v \chi \rceil$ in Eq. \eqref{timelengthdiscreteb}; care must be taken with this relation though since it can skip some values of $n$; for these skipped values, one should use Eq. \eqref{length_contraction}. We now see that for the shortest possible spatial extent in RF2, namely, $\Delta l' = \chi$, one uses $\gamma(v,\Delta t'=\tau)$ and $v=c$ in Eq. \eqref{timelengthdiscreteb}.  But since $\gamma(v,\tau)=1$ regardless of the relative velocities of the two RFs, a $\Delta l'$ of $\chi$ in RF2 is measured as being $\chi$ by observers in RF1. Thus, no length contraction occurs and the immutability of the atom of space is conserved.

We can now revisit Bergson and his view that we all experience his \textit{Time}, with time durations that do not experience dilation.  The results of this section show that only very short durations, on the order of a few integer multiples of $\tau$, do not experience dilations.  However, all important chemical, biological, neurological, and physiological effects transpire over much longer time durations -- durations that experience dilations according to Einstein's theory. Thus, we can close the book on this philosophical debate once and for all.  Einstein did indeed find a ``way of not aging'' -- just be fleet of foot.

\section{\label{sec:LightSpeed}Travel at the Speed of Light}
Besides the modifications to length contraction and time dilation, a straight-forward consequence of the new distance formula is that it allows objects to travel at light speed over certain temporal durations.   This should not be surprising, because motion in DST involves a particle undergoing a certain number of sequential spatial jumps, with each jump being of extent $\chi$ over a duration $\tau$ (with $\chi / \tau = c$).  This sequence of jumps is then followed by one or more durations of $\tau$ where the particle does not move.  But how many sequential jumps of $\chi$ can be done, and at what price, in terms of energy provided to system?  To answer these questions, consider the right side of Fig. \ref{fig:Fig11} that shows a right triangle.  The conventional distance formula given by the Pythagorean theorem does not admit a solution where the lengths of the base and hypotenuse are equal.  However, the new distance formula given by Eq. \eqref{LeopoldB} does, as long as the hypotenuse and base are long enough relative to the triangle's height.  To see this, let the velocity of the light-clock be $v=c$ by setting $m=n$ in Eq. \eqref{LeopoldB}.  Next, choose a particular duration in the rest reference frame $\Delta t' = n'\tau$; this sets $h$ to $h=n'\chi$.  Finally, use Eq. \eqref{LeopoldB} to solve for the critical value of $n$ for which $n=m$ is possible for this and all greater integer values:

\begin{equation}\label{ncritical}
  n_{critical} > \frac{1}{2} \left( n'^2-1 \right).
\end{equation}

\noindent For large $n'$, Eq. \eqref{ncritical} yields $n' \approx \sqrt{2n_{critical}}$, this, along with $\gamma_{critical}=n_{critical}/n'$, $\Delta t' = n'\tau$, and $\Delta t= \gamma_{critical}\Delta t'$ yields

\begin{equation}\label{gammacritical}
  \gamma_{critical} \approx \sqrt{\frac{\Delta t}{2\tau}}.
\end{equation}

\noindent  For time periods $\Delta t$ large relative to $\tau$, we can safely use the conventional equation relating kinetic energy ($KE$) of a particle to $\gamma$, namely, $KE=\left( \gamma - 1 \right) mc^2 \approx \gamma m c^2$.  Upon doing so, it is seen that the energy a particle needs such that it can be measured as traveling at a speed $c$ over a particular duration $\Delta t$ that is large relative to $\tau$ is

\begin{equation}\label{KEcritical}
  KE_{critical} \approx \sqrt{\Delta t / 2\tau}mc^2.
\end{equation}

It is important to note that we are not predicting faster-than-light travel, or even travel at the speed of light for any situation normally encountered or even possible given modern-day technology.  Concerning the former, the rule that a particle transits at most one $\chi$ per $\tau$ is built into the model from the very beginning, corresponding to a maximum velocity of $c$.  Concerning the later, let us consider what is necessary to have a measurement of an electron's velocity yield $v=c$.  Imagine that we can fabricate two detectors tips that can precisely determine the time that an electron has passed underneath them, but otherwise not perturb the speed or trajectory of the electron -- for example two nanofabricated atomic force microscope tips.  With currently available nanofabrication techniques, a separation between the tips of $10$ nm can be achieved.  If the electron is traveling at a speed $c$, then $\Delta t = \Delta d/c \approx 0.1$ fs.  Equation \eqref{KEcritical} then yields a value of approximately 5,000,000 TeV.  This value exceeds what is possible with the most powerful existing particle accelerators by a factor of $10^6$, but may be possible with accelerators constructed in the far future.  A more realizable test is described in the next section, namely, analyzing any anomalous motion of black holes and other astronomical bodies or groups of bodies. (\citet{Crouse2016a, Crouse2016b}).

\section{\label{sec:Crystal}Anomalous Particle Motion in the Discrete Space-Time:
	{\fontsize{12pt}{15pt}\selectfont\textnormal{The Use of Quantum Black Holes as Planckscopes}}}

\subsection{\label{subsec:GravityCrystal}The Universe-Wide Gravity Crystal}
\begin{figure}
	\centering\includegraphics[width=8cm]{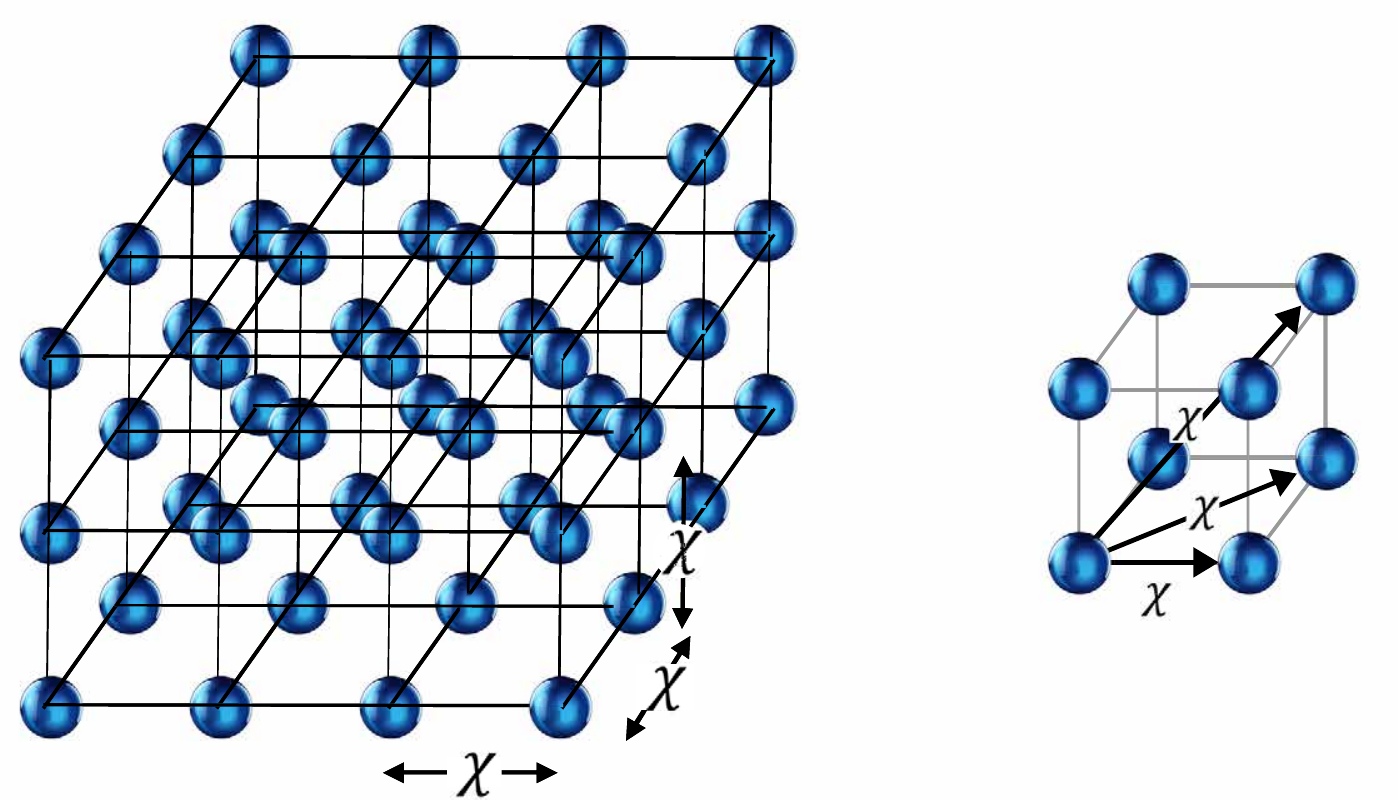}
	\caption{\textbf{Left:} The universe-wide gravity crystal that has a cubic lattice, a lattice constant of $\chi$, and a basis of one particle of mass $m_c$. \textbf{Right:} One unit cell showing that the nearest, next-nearest and next-next-nearest neighbor distances are all the same value $\chi$ -- this is a result of the new distance formula.  This aspect will greatly reduce the anisotropy in energy bands and effective mass observed in Figs. \ref{fig:Fig15} and \ref{fig:Fig16}; the curves in these two figures were calculated using the EPM calculation that itself used the conventional distance formula.}
	\label{fig:Fig14}
\end{figure}

In 1957 John A. Wheeler sought to show that all of classical physics, particle physics included, is ``purely geometrical and based throughout on the most firmly established principles of electromagnetism and general relativity" (\citet{Wheeler1957}).  Wheeler made use of well developed concepts in quantum electrodynamics to demonstrate that the fine structure of space is composed of a random array of quantum ``wormholes'', with each wormhole having a pair of charges, $q_p=\pm\sqrt{4 \pi \epsilon_o \hbar c}$, and each charge having a mass $m=m_p=E / c^2 = \sqrt{\hbar c/G}=2.18 \times 10^{-8}$ kg.  He stated that these charges (i.e., Planck particles) have an average spacing of the Planck length $l_p=1.62 \times 10^{-35}$ m.  Otherwise the particles are randomly distributed; hence he called this system a ``quantum foam''.  However, if space is discretized, a random distribution of Planck particles that includes fractional distances of $\chi$ is not allowed.  Order must be imposed on the structure, changing the foam into a crystal (Fig. \ref{fig:Fig14}).  This structure then forms a gravity crystal (GC) that is described in detail by Crouse in (\citet{Crouse2016a}).  

We include a discussion of particle motion in the GC in this paper because it may provide the best way, perhaps the only realizable way, to experimentally confirm the discrete nature of space and to provide the philosophically important aspect of falsifiability. There have been other ways proposed to verify DST.  Two in particular are the recent papers by Gudder (\citet{Gudder2017}) and Van Bendegem (\citet{VanBendegem2000}).  The work by Gudder used methods from quantum field theory and nicely supplements the earlier work by Crouse in (\citet{Crouse2016a}) that used well established tools in the field of solid-state physics to analyze the GC.  Both works predict similar scattering properties of the GC and anomalous particle motion that mimic effects normally attributed to dark matter and dark energy.  In (\citet{VanBendegem2000}), Van Bendegem considers an interesting construction where the velocity of a particle is proportional to the distance it has traveled and the distance that remains to be traveled, where the particle is confined to travel in the range $x=0 \rightarrow d$, namely, an equation of motion (EoF) of $dx/dt = ax \cdot \left (d-x \right )$.  Van Bendegem describes how the system can be constructed with particular values for $a$ and $d$ such that a particle will tend to either one position (in the limit $t \rightarrow \infty$) if space-time is continuous or chaotically occupy all positions between $0$ and $d$ if space and time are discrete.  However, for this process to satisfy various physical requirements (e.g., the velocity must remain less than $c$), the value of $d$ needs to be extremely small, in the neighborhood of the Planck length. Specifically, it is seen that Van Bendegem's method requires $d$ to be less than $1.56\chi$ in order for chaotic behavior to manifest.  However, $d$ must also be an integer multiple of $\chi$; the only integer multiple of $\chi$ that is less than $1.56\chi$ is $1\chi$. But to be able to construct the system such that it follows, or is governed by the EoF stated above, one needs to be able to work with a system many multiples of $\chi$ in extent, not just a single $\chi$.  Hence, this method cannot be used to verify the discrete nature of space and time \footnote{Matlab code to model Van Bendegem's construction can be obtained from Crouse upon request.}. This result leads us back to the GC in the hope of finding a method to verify and empirically study DST.

The GC is composed of an array of particles, all with identical mass $m_c$, and with one particle at each lattice position $\vec{R}$.  The GC creates a gravitational potential energy profile $V_c$ that is experienced by a particle (electrically neutral and mass $m_{particle}$) traveling within the crystal:

\begin{equation}\label{Coulombic}
V_c(\vec{r})=-G m_{particle} m_c \sum_{\vec{R}} \frac{1}{\left\vert \vec{r}-\vec{R} \right\vert}
\end{equation}

\noindent Any noncrystalline contributions to the potential energy (i.e., $V_{external}$) that may be produced by stars, planets, interstellar gas \dots are added separately to $V_c$ to yield the total potential energy term that is used in Schr\"{o}dinger's equation:

\begin{equation}\label{Schrodinger}
-\frac{\hbar^2}{2 m_{particle}}\nabla^2\psi+\left ( V_c + V_{external} \right)\psi=\mathcal{E}\psi
\end{equation}

Instead of directly solving such a difficult equation (i.e., Eq. \eqref{Schrodinger}), with the large number of terms in $V_c$, we can use the simplifying tools of solid-state physics.  We can do so because both the gravitational and electromagnetic forces have a $1 / r^2$ dependence.

\subsection{\label{subsec:EffectiveMass}The Effective Mass Theory}
The effective mass method allows one to lump all the effects of the crystal particles into one parameter $ m_{inertial}$ that replaces $m_{particle}$ in the kinetic energy term in Schr\"{o}dinger's equation, and then use this term in a much simplified form of the equation for the system (\citet{Ashcroft1976}):

\begin{equation}\label{SchrodingerEffectiveMass}
-\frac{\hbar^2}{2 m_{inertial}}\nabla^2\psi+V_{external}(\vec{r})\psi=E\psi
\end{equation}

To calculate $m_{inertial}$, you first calculate a particle's dispersion curve, which provides the relationship between its energy ($\mathcal{E}$ of Eq. \eqref{Schrodinger}) and wave vector $k$ (i.e., crystal momentum). To calculate this dispersion curve, one can use the tight-binding method (TBM), emperical pseudo-potential method (EPM) or some other well established technique to solve for $\mathcal{E}$ as a function of $k$ (\citet{Ashcroft1976,Vas2017}) \footnote{Matlab code for TBM, EPM and relativistic Kronig-Penney models are available upon request.}.  One then uses $\mathcal{E}$ to calculate $m_{inertial}$ according to:

\begin{equation}\label{effectivemass}
\left(\frac{1}{m_{inertial}}\right)_{i,j} = \frac{1}{\hbar^2} \frac{\partial^2 \mathcal{E}(\vec{k})}{\partial k_i \partial k_j}
\end{equation}

\noindent where $m_{inertial}$ is expressed in its full tensorial form.  Besides $m_{inertial}$, many other interesting properties can be gleaned from a particle's dispersion curve, including the existence of bandgaps and Brillouin zones (BZs).  These things provide important information on how particles behave in crystals, sometimes predicting seemingly bizarre behavior.  For example, energy bandgaps form forbidden energy ranges for particles, but particles can ``jump'' these gaps by acquiring the necessary energy from another particle or excitation.  BZs provide information about the range of momentum that a particle can have, including an effective maximum momentum.  

Let us for the moment consider an analogous system: an electron in silicon.  It is not at all unusual for an electron in silicon to have a negative value for $m_{inertial}$ for particular energies and momenta \footnote{$m_{inertial}$ being different than $m_{gravitational}$ in this situation is not a violation of Einstein's equivalence principal because $m_{inertial}$ should be considered more as a parameter describing motion, due not only to the particle's gravitational mass but also to the complex interactions of the particle with all the crystal particles.}.  A negative value for $m_{inertial}$ indicates that a particle will accelerate in the opposite direction of an applied \textit{external} force (i.e., a force produced by noncrystalline sources, such as an electric field produced by a voltage applied between two electrical contacts in this electron/silicon example). Crystallographers know how to extract information about the structure of a crystal (e.g., the type of lattice and lattice constant) by studying the motion and scattering of particles traveling within the crystal (i.e., using the spectroscopic methods of electron diffraction or X-ray diffraction). We will show in this section that quantum sized black holes (BHs) can play the same role as that played by electrons or X-rays in silicon, namely, as a measurement tool or scope that can be used to glean information about the GC. In other words, quantum BHs can serve, not as a microscope to assess features at the micrometer scale, but rather as a planckscope to study material properties at the Planck scale. 

We now return to the discussion about the GC and John Wheeler.  In (\cite{Wheeler1957}), Wheeler demonstrated that if the constituent crystal particles (again, of charge $q_p=\pm\sqrt{4\pi \epsilon_o \hbar c}$ and mass $m_p=\sqrt{\hbar c / G}$) are separated from each other by an average distance of $l_p$, then the positive mass produced by electromagnetic energy (via $E=mc^2$) is totally compensated by negative mass produced by gravitational energy, such that ``to the extent this compensation holds locally, nearby wormholes exert no gravitational attraction on remote concentrations of mass-energy''.  In (\citet{Crouse2016a}), we considered a GC where this compensation does not happen, therefore the particles that compose the crystal all have a gravitational mass $m_c=m_p$.  We also assumed that the lattice is cubic with lattice constant $\chi=l_p$.  It was seen in (\citet{Crouse2016a}) that a particle traveling within this crystal can exhibit negative and near-zero values for $m_{inertial}$ while its gravitational mass remains a constant $m_p$.  However, no justification was given in (\citet{Crouse2016a}) as to why this compensation does not occur.  In this work however, we have shown that a spacing of $l_p$ is not possible, because $l_p$ is less than the fundamental length $\chi=2l_p$ in our DST model.  If $\chi=2l_p$ is the lattice constant of the GC, then it is easy to show (using Wheeler's methods described in (\citet{Wheeler1957})) that there is an uncompensated mass of $3m_p/8$ for each crystal particle, thus $m_c=3m_p/8$.  Regardless of our particular result for $m_c$, we recognize the fact that there is little agreement within the quantum gravity community on the values of $m_c$ and $\chi$.  Thus, it is useful to develop an equation that predicts which mobile particles will be affected by a GC; we do this next.

While there is some disagreement on the lattice constant of the GC (i.e., $\chi$), the values of $m_c$ advocated by different research groups span a much wider range, from $3.78 \times 10^{-130}$ kg (\citet{Carroll2006}) to $3.78 \times 10^{-8}$ kg (\citet{Milonni2013}).  It is therefore useful to have an equation for the lower limit for the mass of an electrically neutral particle (i.e., $m_{particle}$) such that the particle is affected by the GC.  To develop an estimation of this mass, one equates the kinetic energy term and the dominant potential energy term of $V_c$ in Eq. \eqref{Schrodinger} \footnote{To be more accurate, one could include a degeneracy factor in the right side of Eq. \eqref{massestimateformula} to account for the number of nearest neighbors (e.g., $N=6$ for a cubic lattice).}:

\begin{equation}\label{massestimateformula}
\frac{\hbar^2k^2}{2m_{particle}} = \frac{Gm_{particle}m_c}{\chi}
\end{equation}

\noindent The effects of the crystal most often manifest themselves at the BZ boundary, namely at $k=\pi/\chi$.  Using this value of $k$ in Eq. \eqref{massestimateformula}, one arrives at the following approximation for the mass ($m_{est}$) of a particle that will interact strongly with the GC (again, the GC has a basis of one particle of mass $m_c$, a cubic lattice constant $\chi$, and the mobile particle is electrically neutral):

\begin{equation}\label{massestimate}
m_{est}=\frac{\pi\hbar}{\sqrt{2Gm_c \chi}}
\end{equation}

\begin{figure}[H]
	\centering\includegraphics[width=8.5cm]{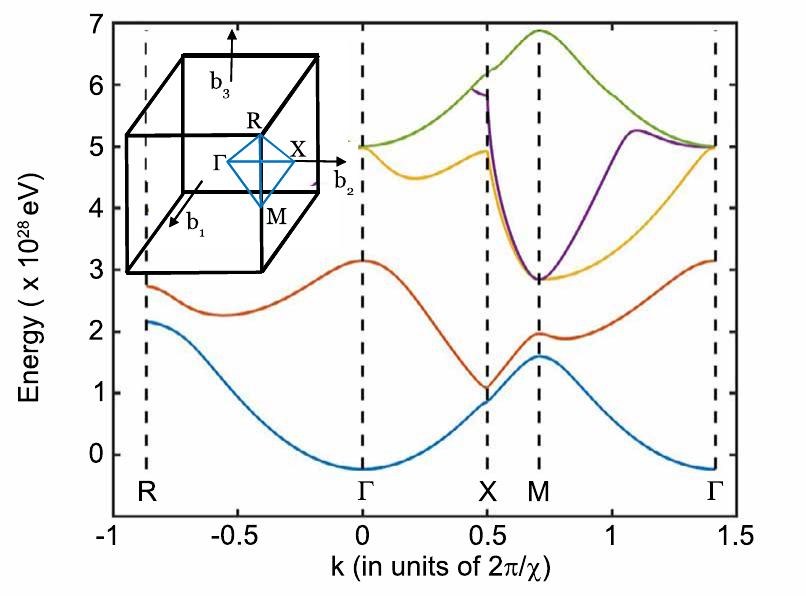}
	\caption{The dispersion curve (calculated using EPM \citet{Crouse2016a}) for a particle with a gravitational mass $m_{particle}=2.59 \times 10^{-8}$ kg traveling within a cubic GC ($m_c=3.78 \times 10^{-8}$ kg and $\chi=2l_p$).  The bands are non parabolic, which is indicative of a $m_{inertial}$ that varies with momentum $k$, as shown in Fig. \ref{fig:Fig16}. \textbf{Inset:} One unit cell of reciprocal space showing the crystal directions.}
	\label{fig:Fig15}
\end{figure}

\begin{figure}[H]
	\centering\includegraphics[width=8.5cm]{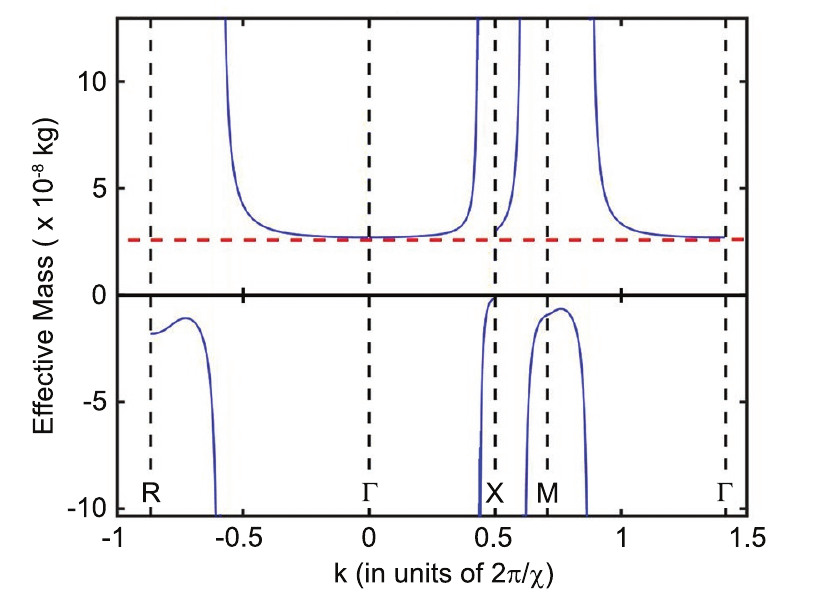}
	\caption{The inertial mass $m_{inertial}$ (blue line) as a function of momentum $k$ of a particle (with gravitational mass $m_{particle}=2.59 \times 10^{-8}$ kg) traveling within the cubic GC ($m_c=3.78 \times 10^{-8}$ kg and $\chi=2l_p$). The red dotted line is the constant gravitational mass $m_{particle}$ and the vertical dotted lines are BZ boundaries. It is seen that away from the $\Gamma$-point, $m_{inertial}$ differs significantly from $m_{particle}$, with $m_{inertial}$ being much greater than $m_{particle}$, near zero, or even negative.}
	\label{fig:Fig16}
\end{figure}

As an example, let us consider two commonly stated values for the vacuum energy density, namely $\xi_1=10^{-9}$ J/$m^3$ (\citet{Carroll2006}) and $\xi_2=10^{113}$ J/$m^3$ (\citet{Milonni2013}), and calculate the corresponding mass $m_c$ (via $m=E/c^2$ with $E=\xi \chi^3$) of each constituent crystal particle assuming a lattice constant of $\chi=2l_p$. For $\xi_1$ we obtain $m_c=3.78 \times 10^{-130}$ kg, and using this value in Eq. \eqref{massestimate} we obtain $m_{est}=2.59 \times 10^{53}$ kg.  In this case, $m_{est}$ is approximately the mass of the entire universe (\citet{Davies2006}), thus no particle with a realistic mass would ever feel the effects of this GC.  For $\xi_2=10^{113}$ J/$m^3$, we obtain $m_c=3.78 \times 10^{-8}$ kg and $m_{est}=2.59 \times 10^{-8}$ kg.  Particles of this gravitational mass are realistic and would be significantly affected by the GC; their energy bands are non parabolic (Fig. \ref{fig:Fig15}) and their values for $m_{inertial}$ can be much different than their gravitational mass, being negative for various ranges of energy and momenta (Fig. \ref{fig:Fig16}).  Again, a negative $m_{inertial}$ indicates that a particle will accelerate in the opposite direction of an external force \footnote{Conservation of energy and momentum are still conserved since the system includes not only the particle but also the entire crystal; the universe-wide GC serves as an near infinite reservoir of energy and momentum.}.  In the case of the universe, the external force is the cumulative gravitational forces due to all planets, stars and galaxies \dots, and is in a direction towards the ``center'' of the universe.  Particles with a negative value of $m_{inertial}$ will be observed to be accelerating in the opposite direction, that is, away from the center of the universe -- these particles will be ``pushed'' by the ``pull'' of gravity.  Such anomalous inertial behavior of astronomical bodies (serving as the probes in our planckscopes) should be easily detectable and quantified using the latest telescopes.  

The results of this section also indicate that black holes (BHs) are more complicated than widely believed.  For a bit of background, any current textbook on general relativity states that BHs have only three properties:  total mass, spin, and electric charge.  However, the results in this paper predict that the distribution of the mass within the event horizon is very important in determining the BH's motion in response to external gravitational forces.  Consider two cases involving BHs with identical total masses traveling within the GC (with $m_c=3.78 \times 10^{-8}$ kg and a lattice constant of $\chi=2l_p$) -- Case 1: a typical stellar black hole of $10M_{Sun}$ where all the mass is concentrated into the ``singularity'' of volume $\chi^3$; Case 2: a BH of the same mass as Case 1, but where the BH is composed of a uniform distribution of particles over the BH's volume of $\left(4 \pi/3 \right) R_s^3$, with $R_s$ being the BH's Schwartzchild radius and equal to $R_s=2G(10M_{Sun})/c^2=2.95 \times 10^{4}$ m, a value that is thirty-nine orders of magnitude greater than $\chi$.  For Case 1, the particle is an elementary particle with a mass much greater than that provided by Eq. \eqref{massestimate}, i.e., $10M_{Sun} \gg 2.59 \times 10^{-8}$ kg.  Hence, the particle will experience a strong gravitational binding to a neighboring crystalline particle at a particular lattice site \footnote{This result is also borne out of the TBM algorithm, which shows $m_{inertial}$ is extremely large for this BH, much larger than its gravitational mass of $10M_{Sun}$. We do not show the plot of $m_{inertial}$, but the reader can verify this using the available TBM Matlab code.}$^,$\footnote{Actually, such a massive particle would more than likely destroy the crystalline order in the neighborhood of the mobile particle. Less massive particles should really be considered for Case 1, but with masses still significantly greater than $2.59 \times 10^{-8}$ kg.}.  Thus, the particle is tightly bound to this particle and will be largely unresponsive to external gravitational forces produced by stars, galaxies and all other non-GC entities.  For Case 2, the mass per $\chi^3$ of volume is $6.77 \times 10^{-87}$ kg -- much less than $2.59 \times 10^{-8}$ kg.  Therefore, this BH will respond to gravity in the expected way, namely with $m_{inertial} \approx m_{gravitational}$. Hence, these two BHs of the same total mass, will be observed to act very differently in response to gravitational forces.

In the calculations of this section, we have used the property of discretization of space to justify the ordering of Wheeler's foam into a crystal, but have used the conventional Pythagorean theorem to calculate distances.  Future work is needed to implement Leopold's theorem into EPM, TBM and other band diagram algorithms.  Because the new distance formula predicts equal distances for a particle's nearest, next-nearest and next-next nearest neighbors (see Fig. \ref{fig:Fig14}), it is reasonable to suspect that the use of Leopold's theorem will significantly reduce the anisotropy of the band diagram and effective inertial mass observed in Figs. \ref{fig:Fig15} and \ref{fig:Fig16}.

\section{\label{sec:Discussion}Discussion}

In the prior sections we focused on straightforward consequences of DST and Eq. \eqref{LeopoldB}; in this section we discuss some remaining topics, open questions, speculate on some issues, and discuss some interesting paths for future study.  First, no work in DST is complete without a discussion on causality.  We then discuss what we think is the sole remaining issue with DST, namely, conservation of energy-momentum.  We discuss how this issue, now not muddied by the other superficial problems, is both glaring and extraordinarily interesting. Next, we discuss possible ways to work gravity into our DST model.  And lastly, we briefly discuss the impact of our DST model on casual set theory.

\subsection{Causality in Discrete Space-Time}
The issue of violations of causality in DST has been debated for hundreds of years by a multitude of philosophers, mathematicians and scientists -- see \citep[76-81]{Hagar2014} for a thorough review of the debate on this issue.  In our view, the important point that has been missed in this debate is that either side of an ``atom'' of space (in fact the entire volume and surface of an atom) is the \textit{same point in real space (i.e., discrete space)} -- one only encounters apparent causality problems when incorrectly viewing the situation from the artificial perspective of continuous space.  Because of this, inquiring about displacements, positions, mechanics, kinematics or anything else \textit{within} any one discrete point (i.e., within any atom of space) is meaningless. Moot then, is the debate as to how a force is instantaneously transmitted across one Weyl-tile (or across a sphere of diameter $\chi$) such that both sides accelerate identically and synchronously in response to a force \citep[76-81]{Hagar2014}. 

\subsection{Conservation of Energy-Momentum, Mach's Principle and the Principle of the Constancy of the Speed of Light}
With the simple problems associated with DST being solved, what emerges is a clearer picture of the challenges (or opportunities) that remain for DST.  We see the sole remaining problem, or issue, as being the conservation of energy-momentum in DST. This issue is rich with implications on motion, inertia and mass.  The problem can be described as follows: \textit{over a duration $\tau$}, motion in DST involves one of only two things, a single spatial translation of extent $\chi$ or no translation. Over longer durations, motion involves a series of translations of $\chi$ (at most one translation of $\chi$ per duration $\tau$) followed by one or more durations of $\tau$ during which the particle does not translate. The velocity $v$ of a particle is the average number of translations (multiplied by $\chi$) over a duration large relative to $\tau$. However, this leads to the obvious question: in this model of motion, for the durations when the particle does not translate, where did its energy and momentum go?  It would appear that conservation of both energy and momentum are regularly violated in the DST model!  Forrest suggests that we reconsider the very nature of momentum.  He proposes that ``momentum need not be defined as mass times velocity, but rather should be understood as a measure of the \textit{tendency} of a particle to move'' and that a ``particle either stays still or moves along but its propensity to move could be constant'' \citep[337-338]{Forrest1995}.  In causal set theory, Henson calls this phenomenon ``swerving away from the geodesic'' (\citet{Henson2008}).  We on the other hand, prefer to look at this as an opportunity to reconsider Mach's principle, and the nature and origin of inertia and inertial mass.  If one accepts the concept that motion happens in this ``jerky'' way \citep[142-143]{VanBendegem1995}, then inertia, inertial mass, energy and momentum must be properties that are emergent at macro spatio-temporal scales.  This is very interesting, because it suggests that inertial mass for a time duration of $\tau$ is zero, a true violation of Einstein's equivalence principle, and that the laws of conservation of energy and momentum are not strictly conserved at these small time durations.  Thus, one is inexorably led to Ernst Mach, D.W. Sciama and George Berkeley and their views on inertia \citep[16]{Sciama1969} (\citet{Ghosh2000}).  And while Mach's principle has been much maligned, I see only artificial problems that are inhibiting its further study and acceptance, similar to the state of DST before this paper. But including the solutions to the problems with Mach's principle is out of the scope of this paper and will need to be a focus of a future paper \footnote{One of the main problems commonly cited is that there is no observed velocity dependence of inertial mass, something contrary to what calculations using Mach's principle ostensibly predict.  However, one should never expect a velocity dependence.  A simple application of the relativity and cosmological principles eliminates this term, as will be shown in a future paper.}$^,$\footnote{A zero inertial mass for time periods $\tau$ makes sense in the context of Mach's principle or the prevailing view of high energy physics.  In both cases, for a time duration $\tau$, there may not be enough time for the particle to interact with the gravitational field (Mach's principle) or the Higgs field (prevailing view); it is these interactions that imbue a particle with inertial mass.}.

One other aspect of our DST model that is worthy of note is that it demotes Einstein's postulate of the constancy of the speed of light to being a consequence of the discretization of space and time, non absolute space, and the concept of motion in our model.  Speed (i.e., velocity) is the ratio of two more fundamental quantities: $\Delta x$ and $\Delta t$.  It is more philosophically sound to postulate the constancy of the foundational components of speed, namely $\chi$ and $\tau$, and have the constancy of $c$ being a mere consequence (i.e., $c=\chi / \tau$).

\subsection{Gravity's Effect on Discrete Space-Time}
One can imagine straightforward ways to work gravity into our model.  In this paper, we will only briefly describe (or speculate on) a couple of ways.  One way is to simply modify Eq. \eqref{LeopoldB} in an obvious way, leading to the following equation for the distance between points $P$ and $Q$:

\begin{equation}\label{gravity}
m \chi_i > \int_{P}^{Q}\sqrt{g_{\mu \nu}dx^\mu x^\nu} - \chi
\end{equation}

\noindent where the repeated indices span the spatial coordinates, and $m$ is the smallest integer that satisfies Eq. \eqref{gravity}. This is very similar to the equation proposed by Arthur March in 1936 (\cite{March1936}).  March himself used a similar method to the one used in this work, namely, based on the ability to measure. Hagar \citep[101]{Hagar2014} calls the approach an ``operationalist theory'', but it is basically just the same logical positivism used in this work.  March's equation is:

\begin{equation}\label{March}
s = \int_{P}^{Q}\sqrt{g_{\mu \nu}dx^\mu x^\nu} - \chi
\end{equation}

\noindent The left sides of Eqs. \eqref{gravity} and \eqref{March} are different: March allows for fractional multiples of $\chi$ while Eq. \eqref{gravity} requires an integer multiple of $\chi$. However, both Eq. \eqref{gravity} and \eqref{March} incorrectly assume an \textit{a priori} existence of space through their use of $g_{\mu \nu}$, something not done in the rest of our work in this paper.  To correctly use LP (or operationalist theory) when gravity is present, one should start by studying gravity's affect on the probes $P_A$ and $P_B$ in Figs. \ref{fig:Fig6} and \ref{fig:Fig7}, as described below.  Again, we are just speculating in this section, a more detailed analysis is outside the scope of this paper. 

To include gravity in DST in a way consistent with LP, one can consider a simple modification to the approach used in Section \ref{sec:SpaceTimeAtom}.  The modification involves assessing the effect of any excess mass ($\Delta$) at a location that is not part of the probe particle. This excess mass (due to the energy density of the gravitational field via $E=\Delta c^2$ or via its connection to the stress-energy tensor) can be considered a parasitic mass because it does not contribute to the localization of the probe particle as calculated from the Compton wavelength $\lambda_c$ but does add to the mass in the equation for $R_S$, the resulting equation for the optimal mass of the probe ($m$) becomes:   

\begin{equation}
\frac{\hbar}{mc}=2\frac{2G \left ( m + \Delta \right )}{c^2}
\end{equation}

\noindent The net result is that the mass of the probe needs to be somewhat smaller than it is in the absence of gravity (denoted below as $m_o$) such that the signal photon ($P_S$) can escape \textit{both} the probe and the excess mass.  With the mass of the probe being smaller, it cannot be as localized as it can be in the absence of gravity. Denoting $m_o$ and $\chi_o$ as the values of the mass of the probe and the hodon respectively in the absence of gravity, the equations for the probe mass and hodon as functions of excess mass ($\Delta$) are:

\begin{subequations} \label{gravityconstants}
\begin{align}
m \left ( \Delta \right ) &=\frac{1}{2}\sqrt{\Delta^2+\frac{\hbar c}{G}}-\frac{\Delta}{2} \approx m_o-\frac{\Delta}{2} \label{GCa}  \\
\chi \left ( \Delta \right ) &= \frac{\hbar}{mc} = \frac{\hbar}{\left ( m_o-\Delta / 2 \right )c} \approx \left ( 1+\frac{\Delta}{2m_o}\right ) \chi_o \label{GCb}
\end{align}
\end{subequations}

We realize that up till this point in the paper we have been arguing that the atom of space should be considered as a constant of nature, and now we are saying that it can vary from one position to another depending on the mass at the positions.  However, this seems to be the most promising way to include gravity into this DST model without sacrificing its philosophically attractive aspects involving non-absolute space and logical positivism.  

Equation \eqref{GCb} shows the atoms of space may vary in size along any path, depending on the excess mass along the path.  Thus the distance ($d$) along any path, where the index $i$ denotes sequential hodons along the path, is:

\begin{equation}\label{gravitydistance}
d=\sum_{i \in Path} \chi_i \left ( \Delta \right )
\end{equation}

\noindent The distance formula, i.e., a version of Eq. \eqref{LeopoldB} or Eq. \eqref{Leopold} but now with gravity present, can be derived in a straightforward way using Eq. \eqref{gravitydistance} in the calculations for the lengths of the base, height and hypotenuse of the triangle. 

Equations \eqref{GCa} and \eqref{GCb} is just a starting point, and they raise many questions.  One question is: how is $\tau$ affected by excess mass and gravitational fields? Is the mass-dependent value of the chronon simply $\tau(\Delta)=\chi(\Delta)/c$? An answer to this and other questions we leave to a future paper.  

\subsection{DST and Causal Set Theory}

Of all the existing approaches to quantum gravity (QG), our DST model is most closely associated with causal set theory (CST). However, the main problem with all existing QC theories, CST included, is that they all strive to ensure Lorentz invariance using the standard forms of the Lorentz transformations, relativistic velocity boosts of RFs, and the other universally accepted but faulty equations of \textit{conventional} SR. Another problem is that most if not all of these QG models also assume the \textit{a priori} existence of a space-time manifold. CST is slightly different in this regard, in that it posits that the ordered causal set (i.e., the causet) is the fundamental structure, not the associated continuum manifold.  Two important changes in thinking are involved in CST: continuum manifolds are replaced by discrete causets, and volume calculations are replaced by the counting of elements (i.e., points or elements).  From a philosophical perspective, the four foundational properties of CST are sound: transitivity, non-circularity, finitarity, and reflexivity (\citet{Reid1999}).  Unfortunately, after this auspicious start, rigorous application of sound philosophic methods and principles are not strictly followed; some questionable assumptions and overly complicated methods are invoked.  Aspects of some of the tools that have been developed in CST will indeed prove to be useful \footnote{Such as the process of determining a manifold from a causet; CST generally does the reverse -- the causet is determined from the manifold via a random sprinkling using a Poisson process.}. However some of the tools have flaws, especially those involving the calculation of time-like and space-like distances as well as issues concerning local Lorentz invariance (LLI). 

Concerning LLI, CST researchers rightly consider their approach to be superior to other QG theories because how it achieves LLI can be more convincingly explained than it can with other theories (\citet{Henson2008}).  CST researchers are also most likely correct in stating that any lattice-based quantum gravity will not have the property of LLI.  However, even in CST, the explanation of LLI resulting from the random nature of the sprinkling of causal elements (using a Poisson process) leaves one a bit unsatisfied (\citet{Henson2008}).  \textit{Our DST model is inherently locally and globally Lorentz invariant} -- it is not a lattice based model and assumes a non absolute nature of space.  This results in no preferred direction and no preferred RF or velocity.

Besides the issues with LLI, the methods used to calculate distances in CST need to be further developed.  There are two accepted methods to calculate time-like distances in CST.  One method involves calculating volumes of hypercones that enclose causal points and then computing the proper time (i.e., time-like distance) between two points (i.e., events) $w$ and $z$ from the volume formula (\citet{tHooft1979}).  A second method is to define the ``proper time $d \left(w,z \right )$, between two related elements $w \prec z$ to be the number of links $L$ in the \textit{longest} chain between (and including) $w$ and $z$, yielding $d(x,y):=L$'' (\citet{Rideout2009}).  The distance calculated in this way is Lorentzian distance, ``in Euclidean spaces...one generally defines distance in terms of shortest paths'' (\citet{Rideout2009}).  Typically in CST, a space-like distance $d \left (x,y \right ) $ is ``given by the minimum time-like distance between an element $w$ in their common past and another $z$ in their common future'' (\citet{Rideout2009}).  Thus, $w \prec \left (x,y \right ) \prec z$, and 

\begin{equation}\label{CSTdistance}
d (x,y)=\underset{w,z}{\text{min}} \, d(w,z)
\end{equation}

\noindent where the time-like distance $d \left ( w,z \right )$ is calculated ``using either the length of the longest chain or volume distance'' (\citet{Rideout2009}).  It is out of the scope of this paper to instruct the reader in greater detail on how to calculate time-like and space-like distances in CST; the important point we want to convey is that in both, counting or determining longest or shortest chains need to be done -- our work shows how to do this accurately.

To conclude this discussion on CST, we reiterate that we believe that CST is a very promising approach to quantum gravity, and that our DST model addresses some of the shortcomings of, and errors in, the methods currently used in CST.  It provides tools to accurately calculate time-like and especially space-like distances. It shows that the standard formula used by CST researchers (e.g., \citet{Myrheim1978}) and others to express the interval between to events in flat space-time, namely $\Delta s^2=c^2\Delta t^2-\Delta x^2-\Delta y^2-\Delta z^2$ is not correct for spatio-temporal scales near the Planck scale. It also shows that CST researchers should not require adherence to the TI theorem (\citet{Brightwell1991}).  It also allows one to calculate a new set of Lorentz transformation equations that are used to assess local and global Lorentz invariance.  Hence, CST, quantum loop gravity, spin-foam approaches, dynamical triangulation methods should all make use of our DST model's method of calculating distances, the new Lorentz factor, and the model's use of non-absolute space.

\section{\label{sec:Conclusion}Conclusion}
It was shown in this work that the DST model requires a new distance formula.  We showed how formulas proposed by Van Bendegem and Forrest could be modified to yield a distance formula that matches one developed by Crouse that is valid at all size-scales.  Our derivation uses the precepts of a restrictive form of logical positivism and assumes non-absolute space that inherently conserves isotropy.  It was shown that when applied to distances near the Planck scale, the new formula yields distances much different than those predicted by the Pythagorean theorem. But for larger length scales, the distances calculated with the new formula converge to those calculated using the Pythagorean theorem.  When using the new distance formula in the otherwise typical derivations of time dilation and length contraction, one sees that the atom of space and atom of time are indeed immutable - true constants of nature and independent of the speed of any observer.  It was also discussed how this new distance formula allows for temporary travel at the speed of light, and how particular empirical tests and observations can be performed to verify the DST model, in particular, the observation and analysis of anomalous motion of astronomical bodies.  We also suggested ways to include gravity into the model.  The main conclusion of the paper is that there are no real problems with DST that cause it to be inferior to the continuous space-time model. Importantly, the proposed DST model yields a new distance formula that should be used in mathematical and physical theories at the Planck scale, including work in the fields of string theory, loop quantum gravity, and causal set theory.  And finally, the model opens up a whole host of new and interesting avenues for investigation/exploration in the fields of mathematics, cosmology, physics and philosophy.

\subsection*{Acknowledgments}
We thank Jean Paul Van Bendegem for discussions about his work on the subject and the encouragement he gave us to write this paper.  I (David Crouse) also wish to thank Leopold Michael Crouse for our numerous discussions on space and time and for his insights he shared with me on the impact of discrete space-time on the Lorentz factor.

\section{Appendix}
Velocity in DST never has a single particular value when expressed in continuous space-time (CST) terms (e.g., $v=0.5c$); velocity is defined only to within an interval in CST terms.  To see this, note that the velocity of a particle is determined by two spatial measurements over a time period.  Let us take the starting position of the particle (in DST terms) as $\tilde{x}_{initial} = s_i\chi$ and the position of the particle after a time $\tilde{t}=r\tau$ (i.e., $r$ ticks of the clock) as $\tilde{x}_{final} = s_f\chi$, where $s_i$, $s_f$ and $r$ are integers.  The velocity of the particle in DST terms is:

\begin{equation}\label{velocity_mapping}
\tilde{v}(s_f,s_i;r)=\frac{\left ( s_f-s_i \right )\chi}{r\tau}=\frac{\left ( s_f-s_i \right )}{r}c
\end{equation}

\noindent with $r \geq 1$, and the maximum of the term $(s_f-s_i)$ being $r$ which represents the maximum of one $\chi$ spatial translation for every $\tau$ temporal duration. Each single velocity value given in DST terms (i.e., Eq. \eqref{velocity_mapping}) is an interval containing a continuous set of velocity values express in CST terms. These intervals are easily constructed by considering the minimum and maximum distances that a particle can travel (in CST terms) for $(s_f-s_i)$ translations occurring over $r$ ticks, and by considering when during the initial tick that the particle has made the transition (i.e., at the start or end of the $\tau$ duration of the tick) and when during the final tick the position was recorded.

\begin{equation}
\tilde{v}(s_f-s_i;r) = 
\left [v_{min},v_{max} \right ]
\end{equation} 

\noindent with

\begin{align}
v_{min} &= 
\begin{cases}
-\frac{c}{r-1},& \hspace{60pt} \text{ if } s_f=s_i\\
\frac{s_f-s_i-1}{r+1}c,& \hspace{60pt} \text{ if } s_f>s_i
\end{cases}\label{v_min_cst} \\
v_{max} &= 
\begin{cases}
\frac{c}{r-1},& \hspace{60pt} \text{ if } s_f=s_i\\
\frac{s_f-s_i+1}{r-1}c,& \hspace{60pt} \text{ if } s_f>s_i
\end{cases}\label{v_max_cst}
\end{align}

\noindent One could say that there is inherent uncertainty in a particle's velocity:  when we say that a particle has traveled from $s_f \chi$ to $s_i \chi$ over a duration $r \tau$, we really only know that the velocity is between $v_{min}$ and $v_{max}$ in CST terms.  If the reader is uncomfortable with CST velocities values less than $-c$ or greater than $c$, which happens with Eqs. \eqref{v_min_cst} and \eqref{v_max_cst}, then you can artificially place an lower bound and upper bound of $-c$ and $c$ on $v_{min}$ and $v_{max}$ respectively without it impacting this model. This is because, again, these CST values for velocities are artificial, it is the DST concept of velocity that is real, namely that a particle has made $s_f-s_i$ spatial translations over $r$ ticks of a clock.



\begin{flushright}
	DAVID CROUSE\\
	Professor and Chairman\\ Department of Electrical\\ and Computer Engineering\\
	Clarkson University\\
	United States of America\\
	\href{mailto:dcrouse@clarkson.edu}{dcrouse@clarkson.edu}\\
	\url{www.clarkson.edu/people/david-crouse}
	
	\vspace{.7cm}
	
	JOSEPH SKUFCA\\
	Professor and Chairman\\ Department of Mathematics\\ 
	Clarkson University\\
	United States of America\\
	\href{mailto:jskufca@clarkson.edu}{jskufca@clarkson.edu}\\
	\url{www.clarkson.edu/people/joseph-skufca}
	
\end{flushright}

\end{document}